\documentclass[prx,twocolumn,amsmath,amssymb,amsthm,superscriptaddress]{revtex4-2}
\usepackage{graphicx,amsmath,relsize,epstopdf,upgreek,color,mathtools,bm,times}
\usepackage[colorlinks=true,linkcolor=blue,citecolor=blue,urlcolor =blue]{hyperref}
\usepackage[hyphenbreaks]{breakurl}
\usepackage{thmtools}
\usepackage{thm-restate}

\newcommand{\norm}[1]{\|#1\|}
\newcommand{\Haar}[1]{\left<#1\right>_\mathrm{Haar}}
\newcommand{\ket}[1]{|{#1}\rangle}
\newcommand{\bra}[1]{\langle{#1}|}

\newcommand{\opinner}[3]{\langle #1|#2|#3\rangle}
\newcommand{\onevec}{\bm{\mathit{1}}}
\newcommand{\rvec}[1]{\pmb{#1}}
\newcommand{\dyadic}[1]{{\bf#1}}
\newcommand{\tr}[1]{\mathrm{tr}\!\left\{#1\right\}}

\newcommand{\D}{\mathrm{d}}
\newcommand{\I}{\mathrm{i}}

\newcommand{\E}[1]{\mathrm{e}^{\mbox{\footnotesize$#1$}}}
\newcommand{\sinc}{\mathrm{sinc}}

\newcommand{\RE}[1]{\mathrm{Re}\!\left\{#1\right\}}

\newcommand{\VAR}[2]{\mathrm{Var}_{#1}\!\left[#2\right]}
\newcommand{\MEAN}[2]{\left<{#1}\right>_{#2}}
\newcommand{\AVG}[2]{\mathbb{E}_{#1}\!\left[#2\right]}

\newcommand{\Uenc}[1]{V_{#1}}
\newcommand{\MSE}[2]{\mathrm{MSE}_{#1}(#2)}
\newcommand{\HARM}[1]{\mathrm{H}_{#1}}
\newcommand{\poly}{\mathrm{poly}}

\begin{document}

\title{Optimized numerical gradient and Hessian estimation for variational quantum algorithms}

\author{Y.~S.~Teo}
\affiliation{Department of Physics and Astronomy, 
	Seoul National University, 08826 Seoul, South Korea}

\begin{abstract}
  Sampling noisy intermediate-scale quantum devices is a fundamental step that converts coherent quantum-circuit outputs to measurement data for running variational quantum algorithms that utilize gradient and Hessian methods in cost-function optimization tasks. This step, however, introduces estimation errors in the resulting gradient or Hessian computations. To minimize these errors, we discuss tunable numerical estimators, which are the finite-difference (including their generalized versions) and \emph{scaled} parameter-shift estimators~[introduced in Phys.~Rev.~A~{\bf103}, 012405~(2021)], and propose operational circuit-averaged methods to optimize them. We show that these optimized numerical estimators offer estimation errors that drop exponentially with the number of circuit qubits for a given sampling-copy number, revealing a direct compatibility with the barren-plateau phenomenon. In particular, there exists a critical sampling-copy number below which an optimized difference estimator gives a smaller average estimation error in contrast to the standard (analytical) parameter-shift estimator, which exactly computes gradient and Hessian components. Moreover, this critical number grows exponentially with the circuit-qubit number. Finally, by forsaking analyticity, we demonstrate that the scaled parameter-shift estimators beat the standard unscaled ones in estimation accuracy under \emph{any} situation, with comparable performances to those of the difference estimators within significant copy-number ranges, and are the best ones if larger copy numbers are affordable.
\end{abstract}

\maketitle

\section{Introduction}

With the inception of quantum information theory~\cite{Chuang:2000fk}, quantum computers and devices~\cite{Ladd:2010aa,Campbell:2017aa,Lekitsch:2017aa} that function according to the laws of quantum mechanics have been envisioned to be the new-age tools for performing computations and other information processing tasks. The subsequent identification of universal gate sets~\cite{Deutsch:1995universality,Barenco:1995elementary,Englert:2001universal,Bartlett:2002efficient,Sawicki:2022universality} motivated many innovative proposals for quantum-computation and cryptographic algorithms~\cite{Grover:1996fast,Shor:1997polynomial,Raussendorf:2001one-way,Kitaev:2003fault-tolerant,Raussendorf:2007topological,Sehrawat:2011test-state,Montanaro:2016quantum}. Despite the theoretical progress, there exist practical challenges that hinder the actual implementation of truly operational quantum devices. These include maintaining the fidelities of qubit sources, unitary gates and measurements~\cite{Knill:1998resilient,Franklin:2004challenges,Aharonov:2008fault-tolerant}, and coping with the large gate complexity needed to construct general-purpose quantum circuits~\cite{knill1995approximation}.

The state-of-the-art in quantum computing technologies revolves around devices that manipulate less than a thousand qubits using noisy unitary gates and measurements---the \emph{noisy intermediate-scale quantum}~(NISQ) devices~\cite{Preskill2018quantumcomputingin}. These devices motivated the development of several kinds of NISQ algorithms~\cite{Bromley:2020applications,Bharti:2022noisy,Finnila:1994quantum,Kadowaki:1998quantum,Aaronson:2011computational,Aaronson:2011linear-optical,Hamilton:2017gaussian,Trabesinger:2012quantum,Georgescu:2014quantum}, of which the class of \emph{variational quantum algorithms}~(VQAs)~\cite{Biamonte:2021universal,Cerezo:2021variational,Cao:2019quantum,Endo:2021hybrid,McArdle:2020quantum} that perform computations in a hybrid manner using both classical and NISQ devices, most commonly discussed in the context of variational quantum eigensolvers designed for quantum-chemistry~\cite{Peruzzo:2014variational,Wecker:2015progress,McClean:2016theory} and combinatorial problems~\cite{Farhi:2014quantum,Zhou:2020quantum}, are of relatively broad interest. In the field of quantum machine learning, VQAs running on circuits that also possess classical-data encodings have also been extensively studied. These include algorithms for classification tasks, nonlinear activation-function implementations, and multivariate function learning tasks~\cite{Schuld:2015introduction,Schuld:2019quantum,Carleo:2019machine,date2020quantum,Perez-Salinas:2020aa,dutta2021singlequbit,Goto:2021universal}. 

Cost-function optimization with VQAs typically requires the statistical sampling of NISQ devices to estimate the gradient and Hessian of the quantum-circuit model function, which are necessary in, for example, steepest gradient-descent~\cite{Boyd:2009dv,Fiurasek:2001mq,Rehacek:2007ml,Teo:2011me} and quantum natural gradient-descent methods~\cite{Amari:1998why,Amari:1998natural,Koczor:2019quantum,Stokes2020quantumnatural,Wierichs:2020avoiding}. Sampling NISQ devices inherently comes with errors originating from statistical fluctuation in the quantum-circuit measurements, which is especially relevant to NISQ devices as currently-achievable noise levels forbid arbitrarily large error-mitigated measurement-data collection within reasonable algorithm runtimes. While the severity of this problem has indeed been raised~\cite{Smart:2019quantum-classical,Endo:2021hybrid} and asymptotic error bounds for sampling the Fisher information in quantum natural gradient methods were derived~\cite{vanStraaten:2021measurement}, more precise error expressions in estimating multiparameter gradients and Hessians on NISQ devices are necessary for developing novel methods that are statistically optimized for VQA executions.

In this article, we examine the estimation accuracies of known methods used to estimate circuit-function gradients and Hessians, namely the (generalized) finite-difference strategy and (scaled)~parameter-shift rule~\cite{Mitarai:2018quantum,Schuld:2019evaluating,Mari:2021estimating}. All of these methods are ``numerical'' except for the \emph{unscaled}~parameter-shift rule, which is ``analytical'': the former approximates gradients and Hessians with a nonzero approximation error, and the latter exactly computes them. We present operational analytical expressions for the \emph{averaged estimation errors} associated with these two kinds of strategies. All expressions are \emph{circuit-averaged} in contrast to those reported in~\cite{Mari:2021estimating}, for instance, which permits the introduction of operationally tunable numerical estimators possessing parameters that can be optimally tuned to minimize NISQ estimation errors. These optimal estimators are designed for a broad class of ``hardware-efficient'' quantum circuits that approximate two-design unitary operators, such as the multilayered \emph{ansatz} comprising single-qubit and controlled-NOT~(CNOT) gates, for which these estimators minimize estimation errors in the initial stages of cost minimization. 

A key observation is that for a given sampling-copy number, the minimized average estimation errors of all numerical estimators scale commensurately with the average gradient- and Hessian-component magnitudes, which in turn drop exponentially with the number of circuit qubits. Without increasing measurement resources, these desirable scaling behaviors prevent the optimally-tuned estimators from effectively making random guesses about the estimated components even in the presence of the barren-plateau phenomenon~\cite{McClean:2018barren,Arrasmith:2021effect,Cerezo:2021cost,Holmes:2022connecting}. Owing to this characteristic, we show that these optimal numerical estimators can outperform those produced by the analytical parameter-shift rule, which does not possess such a characteristic. One striking consequence is that all regimes of sampling-copy numbers in which the optimal (generalized) finite-difference strategies beat the analytical strategy grow exponentially with the circuit-qubit number. 

Last, but not the least, we demonstrate that when one forgoes analyticity and, instead, employ the scaled parameter-shift rule~\cite{Mari:2021estimating}, which is yet another numerical strategy, we find that its estimation accuracy is comparable to those of the numerical difference strategies in orders of magnitude for a certain range of sampling-copy numbers. Beyond this range, the scaled parameter-shift rule exhibits the most favorable estimation accuracy. This further confirms that numerical estimation schemes are better suited for improving gradient and Hessian estimation accuracies as one scales up NISQ devices.

\begin{figure}[t]
	\centering
	\includegraphics[width=\columnwidth]{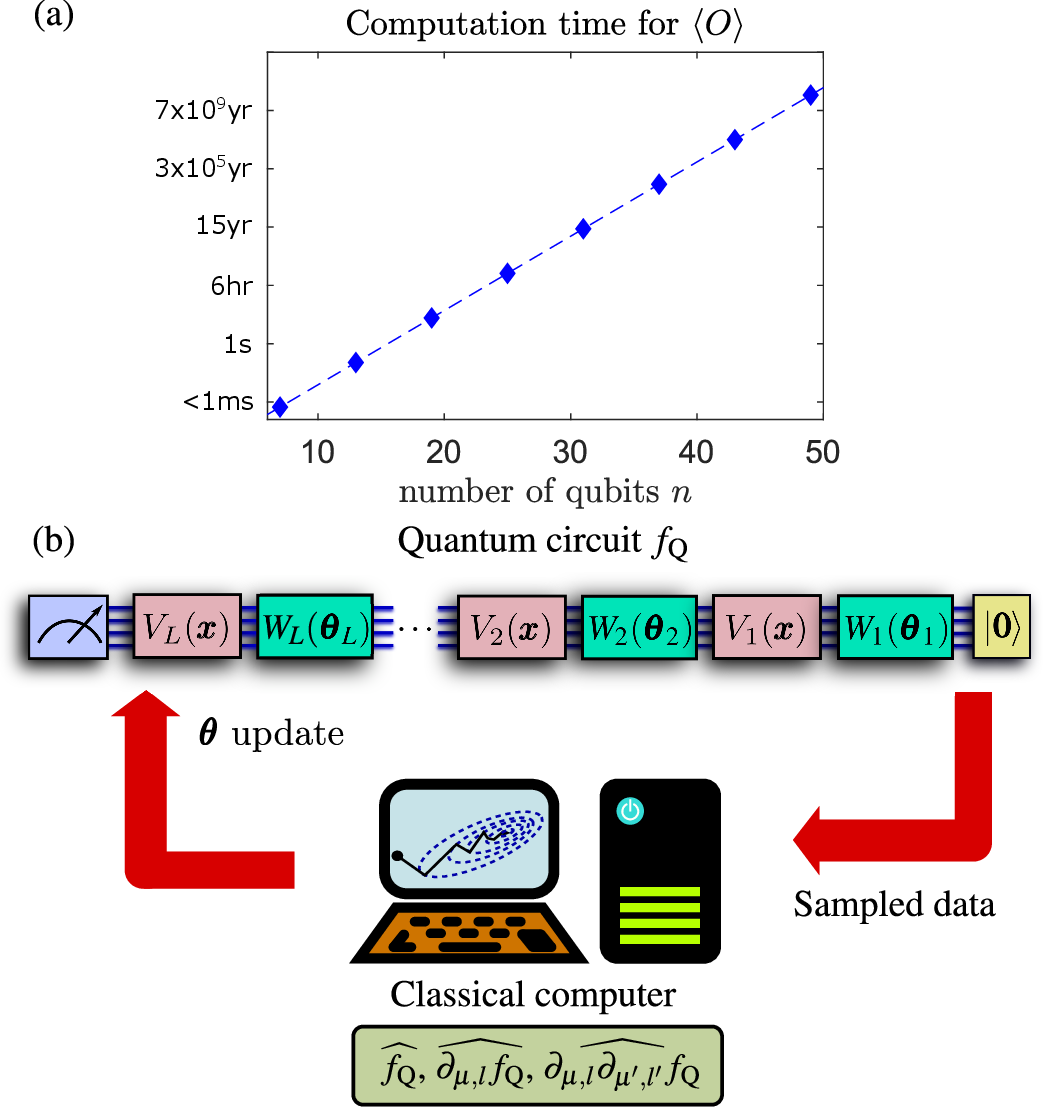}
	\caption{\label{fig:vqa}(a)~Classical computation of $\left<O\right>$ on an AMD Ryzen~9~5900HX CPU. For every $n$, the computation time is averaged over 1000 expectation-value calculations from randomly-chosen $n$-qubit Hermitian operators $O$ and pure states. Owing to memory limitations, a combination of actual computation-time data for observables up to $n=14$ followed by a fitted extrapolation to $n=50$ shows that the average computation time with randomly-generated pure states grows exponentially with $n$. (b)~Schematic of a VQA for minimizing a cost function $C=C[f_\mathrm{Q}]$, which is generally a functional of another parametrized function $f_\mathrm{Q}=f_\mathrm{Q}(\rvec{\theta};\rvec{x})$. Data collected from sampling a NISQ~circuit that models $f_\mathrm{Q}$ using a series of parametrized training $[W_l(\rvec{\theta}_l)]$ and classically-encoded $[V_l(\rvec{x})]$ modules are fed to a classical machine, where the model function $f_\mathrm{Q}$ and its gradient $\partial_{\mu,l}f_\mathrm{Q}$ (and Hessian $\partial_{\mu,l}\partial_{\mu',l'}f_\mathrm{Q}$ if necessary) are estimated for carrying out a pre-chosen optimization scheme in an iterative fashion. Here, the pair $(\mu,l)$ labels the $\mu$th trainable circuit parameter $\theta_{\mu l}$ located in the $l$th trainable module $W_l$. In each step, the updated circuit parameters using the estimated quantities are used to tune the training modules $W_l(\rvec{\theta}_l)$ in the NISQ circuit for subsequent sampling and classical optimization.} 
\end{figure}

\section{Background: Variational quantum algorithms}
\label{sec:vqa}

An especially important and widely-studied computation task is function minimization. In various interdisciplinary applications that are related to quantum mechanics and quantum information, the \emph{cost function}~$C=C[\left<H_j\right>]$ to be minimized is a (real)~functional of expectation values of an (Hermitian) operator set $\{H_j\}$. The expectation $\left<H_j\right>=\tr{\rho H_j}$ is itself a function of a variable state $\rho$ that is optimized in order to attain the minimum value of $C$. An immediate problem with the minimization task is the difficulty in evaluating expectation values of operators describing large physical systems, such as multiqubit systems considered in Fig.~\ref{fig:vqa}(a), using a classical computer. 

\begin{figure}[t]
	\centering
	\includegraphics[width=\columnwidth]{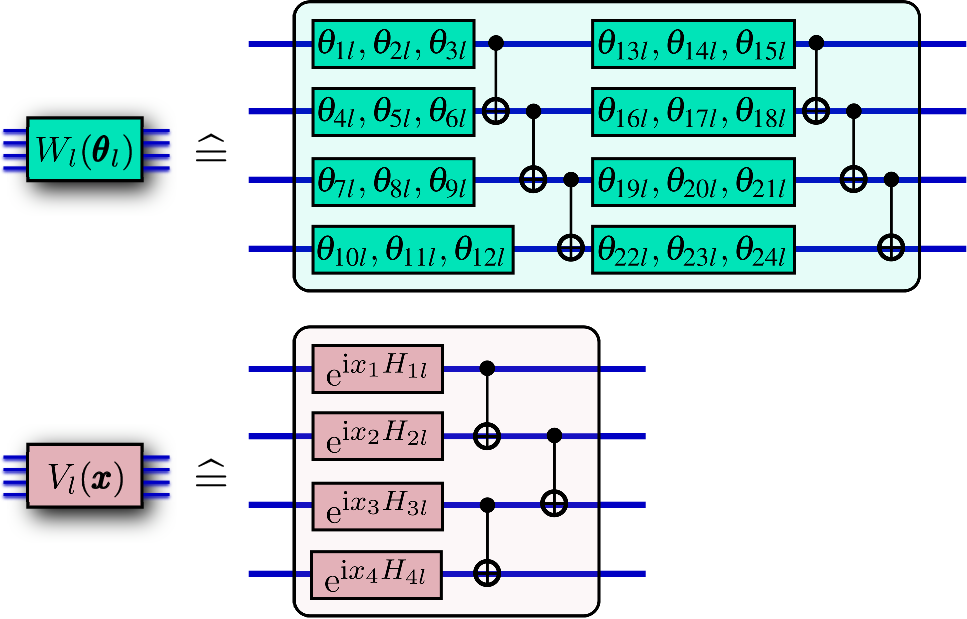}
	\caption{\label{fig:qbitCNOT_ansatz} Examples of ``hardware-efficient'' setups for trainable unitary operators $W_l$ (with two repeated units) and those of fixed encodings $V_l$, shown here for a four-qubit system with $\rvec{\theta}_l=(\theta_{1l}\,\,\theta_{2l}\,\,\ldots\,\,\theta_{24l})^\top$ and $\rvec{x}=(x_1\,\,\ldots\,\,x_4)^\top$. Each single-qubit green block in $W_l$ represents a rotation operator defined by an angle 3-tuple $\phi$. For instance,  $R(\phi=(\theta_{1l},\theta_{2l},\theta_{3l}))=R_Z(\theta_{1l})R_Y(\theta_{2l})R_Z(\theta_{3l})$. The operator $V_l$ may also take similar structures, where $\rvec{x}$ is encoded into entanglement-free unitary operators characterized by their respective generators $H_{kl}$ along with CNOT gates.} 
\end{figure}

In the NISQ era when fully quantum algorithms are out of reach, VQAs are the next viable class of quantum-classical hybrid algorithms for efficient cost-function minimization~[see Fig.~\ref{fig:vqa}(b)]. It makes use of a quantum device, a unitary circuit for instance, to efficiently collect sampled data of expectation values. These data are then transferred to a classical computer that performs an iterative update on the current quantum-circuit parameters using a pre-chosen optimization scheme, which are then used to tune the quantum device for another round of sampling. The action of quantum-circuit tuning is also widely termed quantum-circuit ``training'', borrowing terminologies from machine learning. The entire iterative VQA terminates after the cost function $C$ is minimized. An arbitrary quantum circuit of the NISQ device used to run the VQA is a sequence of training [$W_l(\rvec{\theta}_l)$] and classically-encoded [$V_l(\rvec{x})$] unitary operators, where $\rvec{\theta}_l$ are trainable parameters and $\rvec{x}$ are nontrainable ones. 

The unitary operator $U_{\rvec{\theta};\rvec{x}}\equiv\prod^1_{l=L}V_l(\rvec{x})W_l(\rvec{\theta}_l)$ of a finite depth $L$ represents the most general quantum-circuit model for all VQA applications; here, $\rvec{\theta}$ and $\rvec{x}$ are shorthands for the respective \emph{complete} sets of parameters. If the fixed initial pure product state $\ket{\rvec{0}}\bra{\rvec{0}}=(\ket{0}\bra{0})^{\otimes n}$ is prepared, the relevant cost function $C=C[f_\mathrm{Q}(\rvec{\theta};\rvec{x})]$ is then defined in terms of the circuit model function 
\begin{equation}
	f_\mathrm{Q}(\rvec{\theta};\rvec{x})=\opinner{\rvec{0}}{U_{\rvec{\theta};\rvec{x}}^\dag O\,U_{\rvec{\theta};\rvec{x}}}{\rvec{0}}
	\label{eq:genfQ}
\end{equation}
and measurement observable $O$. As specific examples of VQAs, in \emph{variational quantum eigensolver}~(\emph{VQE}) problems~\cite{Peruzzo:2014variational,Wecker:2015progress,McClean:2016theory} and \emph{quantum approximate optimization algorithms}~(\emph{QAOA})~\cite{Farhi:2014quantum,Zhou:2020quantum}, only trainable operators $W_l(\rvec{\theta}_l)$ are used to minimize the linear cost function $C=\opinner{\rvec{0}}{U_{\rvec{\theta}}^\dag H U_{\rvec{\theta}}}{\rvec{0}}$, where $O=H$ is a Hamilton operator that either describes the dynamics of a physical system, or corresponds to a combinatorial problem. In quantum machine learning tasks~\cite{Schuld:2015introduction,Schuld:2019quantum,Carleo:2019machine,date2020quantum,Perez-Salinas:2020aa,dutta2021singlequbit,Goto:2021universal}, a VQA is employed, for instance, to train the quantum circuit defined by $U_{\rvec{\theta};\rvec{x}}$ to learn a particular multivariate function mapping $f(\rvec{x})$ for different classical-data encoding parameters $\rvec{x}$. Given $m$ of these parameters $\{\rvec{x}_j\}^m_{j=1}$, the quality of the learning procedure is defined by more sophisticated cost functions such as the mean squared-error $C=\sum^m_{j=1}[f_\mathrm{Q}(\rvec{\theta};\rvec{x}_{j})-f(\rvec{x}_j)]^2/m$.

Gradient-based (and Hessian-based) optimization routines in VQAs would then rely on the computation accuracies of $f_\mathrm{Q}$, $\partial_{\mu,l}f_\mathrm{Q}$ and $\partial_{\mu,l}\partial_{\mu',l'}f_\mathrm{Q}$, with all arguments dropped from hereon for notational simplicity unless otherwise necessary. The estimations of gradient $\partial_{\mu,l}f_\mathrm{Q}$ and Hessian components $\partial_{\mu,l}\partial_{\mu',l'}f_\mathrm{Q}$ require the specifications of actual physical \emph{ansatz} structures that make up the $W_l$ and $V_l$ operators. A very common type of parametrized quantum-circuit multiqubit \emph{ans{\"a}tze} encode training parameters $\rvec{\theta}$ on single-qubit rotation unitary operators that are easy to manipulate. They consist of alternating layers of single-qubit rotation and CNOT gates~\cite{Chuang:2000fk,Du:2020aa,Schuld:2020circuit-centric,Schuld:2021aa}, examples of which are illustrated in Fig.~\ref{fig:qbitCNOT_ansatz}. In general, we may further decompose any single-qubit rotation gate $R(\rvec{\phi})$ defined by the trivariate angular parameter $\rvec{\phi}=(\phi_1\,\,\phi_2\,\,\phi_3)^\top$, $R(\rvec{\phi})=R_Z(\phi_1)R_Y(\phi_2)R_Z(\phi_3)$, into a sequence of rotations about the $Y$- and $Z$-axes, so that each basic Pauli rotation gate $R_j(\phi_j)=\E{-\I\phi_j\sigma_j/2}$ is attributed to the relevant single-qubit Pauli operator $\sigma_j$ of the $j$th axis. 

To set the stage for analyzing gradient and Hessian estimations, we shall consider the following ``hardware-efficient'' scenario:
\begin{itemize}
	\item We focus on \emph{Pauli-encoded parametrized quantum circuits}~(PEPQCs). Such a circuit comprises an \emph{arbitrary} chain of trainable ($W_l$) and nontrainable $(V_l)$ encoding modules as shown in Fig.~\ref{fig:vqa}, where all trainable parameters $\rvec{\theta}$ in $W_l$ are encoded onto Pauli rotation operators that constitute single-qubit rotation gates, and CNOT gates shall be used to generate entanglement between all qubits (as in Fig.~\ref{fig:qbitCNOT_ansatz}). These circuits are widely employed in common VQAs (such as quantum eigensolvers and quantum machine learning) and other quantum tasks~\cite{Kandala:2017hardware-efficient,Mitarai:2019generalization,Du:2020aa,Schuld:2020circuit-centric,Schuld:2021aa,Kim:2021hardware-efficient}.
	\item The structures of $V_l$ that house the nontrainable parameters $\rvec{x}$ can be arbitrary.
	\item With no loss of generality, the measurement observable is some traceless Hermitian operator $O=\sum_kh_kO_k/\norm{\rvec{h}}$ written in terms of the multiqubit Pauli basis operators $O_k$. Any Hermitian operator is then $O$ displaced by a multiple of the identity up to normalization. We shall assume that each $O_k$ is sampled \emph{independently} for the same set of PEPQC parameters.
\end{itemize}
We note that the set of $W_l$s, of circuit depths polynomial in the number of qubits $n$, consisting of randomized single-qubit rotation and CNOT gates, are also approximately two-design circuits~\cite{harrow_random_2009}; that is, the operator moments $\left<W_l\right>$ and $\left<W_l^{\otimes2}\right>$ are approximately those from the Haar measure~\cite{Puchala_Z._Symbolic_2017} over the $\mathrm{U}(2^n)$ group.

\section{Figure of merit: mean squared-error}

Throughout the article, separate notations for averages of different kinds are adopted to avoid confusion when several kinds appear at the same time. Averages over (unitary) operator spaces are denoted by the angled parentheses $\MEAN{\vphantom{M}\cdots\vphantom{M}}{}$, which, for instance, could mean averages according to the Haar measure. Numerical or vectorial averages over NISQ-sampling distributions shall be denoted by $\AVG{}{\vphantom{M}\cdots\vphantom{M}}$. Those over the nontrainable parameters $\rvec{x}$ are denoted by $\overline{\vphantom{M}\cdots\vphantom{M}}$.

To explicitly quantify errors for estimating the model function $f_\mathrm{Q}$, its gradient ($\partial_{\mu,l}f_\mathrm{Q}$) and Hessian components ($\partial_{\mu,l}\partial_{\mu',l'}f_\mathrm{Q}$) in any gradient- or Hessian-based methods, we shall examine the mean squared-error~(MSE)
\begin{align}
	\MSE{}{Y}=&\,\overline{\MEAN{\AVG{}{(\widehat{Y}-Y)^2}}{}}\,,
\end{align}
where $\widehat{Y}$ is some generic estimator of the true component $Y$, which can refer to either the quantum-circuit function, its gradient or Hessian component, obtained from sampling the NISQ device. As we shall soon realize, estimators of the latter two can be computed from direct sampling of quantum-circuit functions [defined in Eq.~\eqref{eq:genfQ}] evaluated at various translated circuit parameters. 

For VQAs running on ($d=2^n$)-dimensional circuits, measurements of the circuit observable $O=\sum_kh_kO_k/\norm{\rvec{h}}$ are performed independently with respect to the individual Pauli components $O_k$~(see, for instance,~\cite{Peruzzo:2014variational}). Using the spectral decomposition of the Pauli observable $O_k=\sum^{d-1}_{l=0}\ket{o_{kl}}o_{kl}\bra{o_{kl}}$, sampling the individual function components
\begin{equation}
	f_{\mathrm{Q},k}(\rvec{\theta};\rvec{x})=\sum^{d-1}_{l=0}o_{kl}\left|\opinner{o_{kl}}{U_{\rvec{\theta};\rvec{x}}}{\rvec{0}}\right|^2\equiv\sum^{d-1}_{l=0}o_{kl}p_{kl}(\rvec{\theta};\rvec{x})
	\label{eq:smp_genfQ}
\end{equation}
of $f_\mathrm{Q}=\sum_kh_kf_{\mathrm{Q},k}/\norm{\rvec{h}}$ is equivalent to sampling the circuit probabilities $\sum^{d-1}_{l=0}p_{kl}(\rvec{\theta};\rvec{x})=1$, where the eigenvalues $o_{kl}=\pm1$. Since $\widehat{Y}-Y=\sum_kh_k(\widehat{Y}_k-Y_k)/\norm{\rvec{h}}$, it can be deduced easily, as a result of independence sampling with the $O_k$s, that $\MSE{}{Y}=\sum_kh_k^2\,\MSE{}{Y_k}/\|\rvec{h}\|^2$ if the estimators $\widehat{Y}_k$s are unbiased~($\mathbb{E}[\widehat{Y}_k]=Y_k$).

A usual physical circumstance when sampling a circuit function defined by the measurement observable $O_k$ is when detectors measuring the probabilities $p_{kl}(\rvec{\theta};\rvec{x})$ register ``click''s at different detection sites $l$ one at a time in a randomized sequence, where each ``click'' is statistically independent from the rest, so that a total of $N$ clicks are registered. The resulting sampling distribution is multinomial, where the relative frequencies $\nu_{kl}(\rvec{\theta};\rvec{x})=n_{kl}(\rvec{\theta};\rvec{x})/N\rightarrow p_{kl}(\rvec{\theta};\rvec{x})$ corresponding to the number of ``clicks'' $n_{kl}(\rvec{\theta};\rvec{x})$ registered at the $l$th detector, with $N=\sum^{d-1}_{l=0}n_{kl}(\rvec{\theta};\rvec{x})$. Thus, the average accuracy in estimating $f_\mathrm{Q}$ using the unbiased estimator
\begin{equation}
	\widehat{f}_\mathrm{Q}=\sum_k\dfrac{h_k}{\norm{\rvec{h}}}\widehat{f_{\mathrm{Q},k}}=\sum_k\dfrac{h_k}{\norm{\rvec{h}}}\sum^{d-1}_{l=0}o_{kl}\,\nu_{kl}
	\label{eq:fest}
\end{equation}
is quantified by the MSE.

We caution the reader that the MSEs involve circuit averages of gradient and Hessian components that are generally non-trivial to evaluate. As the present discussion hinges on the two-design approximative PEPQC \emph{ansatz}, explicit MSE expressions are only available when trainable circuits are sufficiently deep for two-design modules to exist. It could very well be that a gradient operation separates a two-design module into two subcircuits, each of which may or may not be deep enough to be a two-design. In most cases, only MSE upper bounds are calculable. While explicit details on how circuit averaging is carried out are supplied in Appendix~\ref{subapp:premise}, we simply state that throughout the main text, unless otherwise stated, all analytical MSE expressions shall refer to the so-called \emph{two-design sandwich}~(TDS) condition ({\bf Case~I} in Fig.~\ref{fig:cases}), where every gradient operation is sandwiched between two two-design trainable modules. Upper-bound expressions for all other cases are obtained from Tab.~\ref{tab:avg_res} in Appendix~\ref{app:QCavg}, and also Appendix~\ref{app:smp_err}.

\section{Results: Sampling errors in gradient and Hessian estimations}

\subsection{Finite (generalized) difference gradient and Hessian methods}
\label{subsec:FD}

The well-known finite-difference~(FD) numerical strategy~\cite{Olver:2014introduction,Guerreschi:2017practical} approximates the gradient components $\partial_{\mu,l}f_{\mathrm{Q},k}$ for each observable basis operator $O_k$ according to
\begin{align}
	\mathrm{FD}^{\epsilon}_{\mu,l}f_{\mathrm{Q},k}=&\,\dfrac{f_{\mathrm{Q},k}(\theta_{\mu l}+\epsilon/2;\rvec{x})-f_{\mathrm{Q},k}(\theta_{\mu l}-\epsilon/2;\rvec{x})}{\epsilon}\nonumber\\
	=&\,\sinc(\epsilon/2)\,\partial_{\mu l}f_{\mathrm{Q},k}
	\label{eq:FD_est1}
\end{align}
by sampling the two quantum-circuit functions \mbox{$f_{\mathrm{Q},k}(\theta_{\mu l}+\epsilon/2;\rvec{x})$} and $f_{\mathrm{Q},k}(\theta_{\mu l}-\epsilon/2;\rvec{x})$, each with trainable parameters displaced by equal magnitudes of $\epsilon/2$ for some $\epsilon>0$. The second equality originates from the usage of Eq.~\eqref{eq:trans}. This particular form approximates $\partial_{\mu,l}f_{\mathrm{Q},k}$ up to $O(\epsilon^2)$ since $\sinc(\epsilon/2)\cong1-\epsilon^2/24$ for a small $\epsilon$. As a consequence of Eq.~\eqref{eq:trans} as well, the Hessian approximator for $\partial_{\mu,l}\partial_{\mu',l'}f_{\mathrm{Q},k}$, defined by applying a second operation $\mathrm{FD}^{\epsilon}_{\mu',l'}$ on $\mathrm{FD}^{\epsilon}_{\mu,l}f_{\mathrm{Q},k}$ in \eqref{eq:FD_est1}, reads 
\begin{align}	
	&\,\mathrm{FD}^{\epsilon}_{\mu,l}\mathrm{FD}^{\epsilon}_{\mu',l'}f_{\mathrm{Q},k}=[\sinc(\epsilon/2)]^2\partial_{\mu,l}\partial_{\mu',l'}f_{\mathrm{Q},k}\nonumber\\
	=&\,\dfrac{f_{\mathrm{Q}}(\theta_{\mu l}+\frac{\epsilon}{2},\theta_{\mu' l'}+\frac{\epsilon}{2};\rvec{x})-f_\mathrm{Q}(\theta_{\mu l}+\frac{\epsilon}{2},\theta_{\mu' l'}-\frac{\epsilon}{2};\rvec{x})}{\epsilon^2}\nonumber\\
	&\,-\dfrac{f_{\mathrm{Q}}(\theta_{\mu l}-\frac{\epsilon}{2},\theta_{\mu' l'}+\frac{\epsilon}{2};\rvec{x})-f_\mathrm{Q}(\theta_{\mu l}-\frac{\epsilon}{2},\theta_{\mu' l'}-\frac{\epsilon}{2};\rvec{x})}{\epsilon^2}\,.
	\label{eq:FD2_est1}
\end{align}
The corresponding function estimators that enter $\widehat{\mathrm{FD}^{\epsilon}_{\mu,l}f_{\mathrm{Q},k}}$ and $\widehat{\mathrm{FD}^{\epsilon}_{\mu,l}\mathrm{FD}^{\epsilon}_{\mu',l'}f_{\mathrm{Q},k}}$ take the form stated in Eq.~\eqref{eq:fest}. Thus, three independent function expectation values are measured for estimating each diagonal~($\mu=\mu'$ and $l=l'$) Hessian approximator component, and four independent function expectation values are measured for each off-diagonal~($\mu\neq\mu'$ and/or $l\neq l'$) component.

The nonzero-$\epsilon$ gradient and Hessian approximators of the FD strategy, collectively denoted by $\widehat{Y}_{k,\epsilon}=\widehat{\mathrm{FD}^{\epsilon}_{\mu,l}f_{\mathrm{Q},k}},\widehat{\mathrm{FD}^{\epsilon}_{\mu,l}\mathrm{FD}^{\epsilon}_{\mu',l'}f_{\mathrm{Q},k}}$, incur errors from \emph{both} finite-copy sampling and nonzero-$\epsilon$ approximation. The corresponding MSE is hence decomposable into these two error types:
\begin{align}
	\MSE{}{Y}=&\,\overline{\MEAN{\AVG{}{\left(\widehat{Y}_{k,\epsilon}-Y_k\right)^2}}{}}\nonumber\\
	=&\,\underbrace{\overline{\MEAN{\AVG{}{\left(\widehat{Y}_{k,\epsilon}-Y_{k,\epsilon}\right)^2}}{}}}_{\displaystyle\text{finite-copy error}}+\underbrace{\overline{\MEAN{\left(Y_{k,\epsilon}-Y_k\right)^2}{}}}_{\displaystyle\text{approx. error}}\nonumber\\
	\equiv&\,\Delta^2_\mathrm{copy}(Y_{k,\epsilon})+\Delta^2_{\epsilon}(Y_k)\,,
\end{align}
where $\Delta^2_{\epsilon=0}(Y_k)=0$, since $Y_{k,\epsilon=0}=Y_k=\partial_{\mu,l}f_{\mathrm{Q},k},\partial_{\mu,l}\partial_{\mu',l'}f_{\mathrm{Q},k}$ according to the definitions in \eqref{eq:FD_est1} and \eqref{eq:FD2_est1}. The finite-copy error $\Delta^2_\mathrm{copy}$ for the gradient and Hessian FD estimators can be directly calculated using \eqref{eq:avgf2} and \eqref{eq:MSE_f} as it only involves linear combinations of squared circuit functions. The nonzero-$\epsilon$ approximation error $\Delta^2_{\epsilon}$, on the other hand, requires the evaluation of $\MEAN{\big(\partial_{\mu,l}f_{\mathrm{Q},k}\big)^2}{}$ and $\MEAN{\big(\partial_{\mu,l}\partial_{\mu',l'}f_{\mathrm{Q},k}\big)^2}{}$ over random circuit parameters. 

With that said, by defining the \emph{total} number of copies $N_\mathrm{T}$ \emph{distributed equally} to all sampled quantum-circuit functions for one FD approximator \emph{per circuit-observable basis operator}~(that is, $N_\mathrm{T}=2N$ for $\widehat{\mathrm{FD}^{\epsilon}_{\mu,l}f_{\mathrm{Q},k}}$, and $N_\mathrm{T}=3N$ and $4N$ respectively for the diagonal and off-diagonal $\widehat{\mathrm{FD}^{\epsilon}_{\mu,l}\mathrm{FD}^{\epsilon}_{\mu',l'}f_{\mathrm{Q},k}}$ components), we list the MSE~formulas for estimating gradient and Hessian components using the FD strategy that is applicable to any PEPQC under the TDS condition:
\begin{align}
	\MSE{\mathrm{FD}}{\partial f_\mathrm{Q}}=&\,\overbrace{\dfrac{4d}{N_\mathrm{T}(d+1)\epsilon^2}}^{\displaystyle\mathclap{\text{finite-copy error}}}+\overbrace{\dfrac{d^2[1-\sinc(\epsilon/2)]^2}{2(d+1)(d^2-1)}}^{\displaystyle\text{approx. error}}\,,\nonumber\\
	\MSE{\mathrm{FD}}{\partial\partial f_\mathrm{Q}}=&\,\dfrac{18d}{N_\mathrm{T}(d+1)\epsilon^4}+\dfrac{d^2\{1-[\sinc(\epsilon/2)]^2\}^2}{2(d+1)(d^2-1)}\,,\nonumber\\
	&\,\qquad\qquad{\text{(diagonal Hessian components)}}\nonumber\\
	\MSE{\mathrm{FD}}{\partial\partial' f_\mathrm{Q}}=&\,\dfrac{16d}{N_\mathrm{T}(d+1)\epsilon^4}+\dfrac{d^4\{1-[\sinc(\epsilon/2)]^2\}^2}{4(d+1)(d^2-1)^2}\,.\nonumber\\
	&\,\qquad\quad\!\!\!{\text{(off-diagonal Hessian components)}}
	\label{eq:FD_MSE}
\end{align}
For optimal estimation of each kind of components using either \eqref{eq:FD_est1} or \eqref{eq:FD2_est1}, the value of $\epsilon$ is therefore chosen such that it minimizes the corresponding operational $\mathrm{MSE}_{\mathrm{FD}}$ in \eqref{eq:FD_MSE}.

We may also consider a type of generalized difference~(GD) estimation strategy~\cite{Swartz:1969generalized} where the corresponding nonzero-$\epsilon$ gradient and Hessian approximators,
\begin{align}
	\mathrm{GDgrad}^{J,\epsilon}_{\mu,l}f_{\mathrm{Q},k}\equiv&\,\sum^J_{j=1}c_j\mathrm{FD}^{j\epsilon}_{\mu,l}f_{\mathrm{Q},k}\,,
	\label{eq:GD1_est}\\
	\mathrm{GDHess}^{J,\epsilon}_{\mu,l;\mu',l'}f_{\mathrm{Q},k}	\equiv&\,\sum^J_{j=1}c_j\mathrm{FD}^{j\epsilon}_{\mu,l}\mathrm{FD}^{j\epsilon}_{\mu',l'}f_{\mathrm{Q},k}\,,
	\label{eq:GD2_est}
\end{align}
are weighted sums of FD approximators of integer multiples of the step size $\epsilon$, which become the true gradient and Hessian components when \mbox{$\epsilon=0$} under the normalization constraint $\sum^J_{j=1}c_j=\rvec{c}^\top\onevec=1$
for the coefficient column $\rvec{c}$, where $\onevec$ is the $J$-dimensional column of ones. This strategy estimates each GDgrad, diagonal and off-diagonal GDHess component by respectively sampling $2J$, $2J+1$ and $4J$ function expectation values, consistent with the number of measurements needed when $J=1$. The forms of their MSE expressions under the TDS condition are rather technical and are instead given in Appendix~\ref{subapp:GD}.

\subsection{Parameter-shift rule}

Both the FD and GD numerical strategies can in general be applied to approximate gradients and Hessians of \emph{any} function that is not restricted to those originating from PEPQCs. A common criticism against these strategies is that the FD and GD approximators require extremely small $\epsilon$ to achieve good approximation qualities. For PEPQCs, using the identity in Eq.~\eqref{eq:trans}, it is indeed easy to verify that, for \emph{any} $s\geq0$,
\begin{align}
	&\!\!\!\!\!\partial_{\mu,l}f_{\mathrm{Q}}=\mathrm{PS}_{\mu,l}f_{\mathrm{Q}}=\dfrac{f_{\mathrm{Q}}(\theta_{\mu l}+s;\rvec{x})-f_{\mathrm{Q}}(\theta_{\mu l}-s;\rvec{x})}{2\,\sin s}\,,
	\label{eq:PS_est3}\\
	&\,\partial_{\mu,l}\partial_{\mu',l'}f_{\mathrm{Q}}=\mathrm{PS}_{\mu,l}\mathrm{PS}_{\mu',l'}f_\mathrm{Q}\nonumber\\
	=&\,\dfrac{f_{\mathrm{Q}}(\theta_{\mu l}+s,\theta_{\mu' l'}+s;\rvec{x})-f_\mathrm{Q}(\theta_{\mu l}+s,\theta_{\mu' l'}-s;\rvec{x})}{4(\sin s)^2}\nonumber\\
	&\,-\dfrac{f_{\mathrm{Q}}(\theta_{\mu l}-s,\theta_{\mu' l'}+s;\rvec{x})-f_\mathrm{Q}(\theta_{\mu l}-s,\theta_{\mu' l'}-s;\rvec{x})}{4(\sin s)^2}\,.
	\label{eq:PS2_est3}
\end{align}
That there exist exact gradient and Hessian expressions for PEPQCs \emph{via} simple training-parameter translations prompted the term ``parameter-shift rule''~(PS)~\cite{Mitarai:2018quantum,Schuld:2019evaluating,Mari:2021estimating}. The corresponding MSEs are therefore just finite-copy errors given by
\begin{align}
	\MSE{\mathrm{PS}}{\partial f_\mathrm{Q}}=&\,\dfrac{d}{N_\mathrm{T}(d+1)(\sin s)^2}\geq\dfrac{d}{N_\mathrm{T}(d+1)}\,,\nonumber\\
	\MSE{\mathrm{PS}}{\partial\partial f_\mathrm{Q}}=&\,\dfrac{9d}{8N_\mathrm{T}(d+1)(\sin s)^4}\geq\dfrac{9d}{8N_\mathrm{T}(d+1)}\,,\nonumber\\
	&\,\qquad\qquad{\text{(diagonal Hessian components)}}\nonumber\\
	\MSE{\mathrm{PS}}{\partial\partial' f_\mathrm{Q}}=&\,\dfrac{d}{N_\mathrm{T}(d+1)(\sin s)^4}\geq\dfrac{d}{N_\mathrm{T}(d+1)}\,.\nonumber\\
	&\,\qquad\quad\!\!\!{\text{(off-diagonal Hessian components)}}
	\label{eq:PS_MSE}
\end{align}
These MSEs are minimized when $s=\pi/2$, which is the \emph{standard} shift value that we shall consider for these analytical estimators.

The analytical PS strategy is now widely accepted as the go-to approach for estimating gradient and Hessian components. An attractive feature is the absence of approximation errors~\mbox{($\Delta_\epsilon^2=0$)}, unlike FD or GD methods. Because of this, it is a belief that FD, for instance, which requires small $\epsilon$ values, would necessitate a large $N_\mathrm{T}$ in order to achieve comparable estimation errors. On the contrary, in Secs.~\ref{sec:optest} and \ref{sec:perf}, we show that there exist exponentially-growing sampling regimes where optimally-tuned FD and GD strategies can achieve very small estimation errors and outperform even the standard PS method.

\subsection{Scaled parameter-shift rule}
\label{sec:scaled_PS}

The standard PS, along with the entire class of parameter-shift rules, form the analytical strategy that exactly computes gradients and Hessians for PEPQCs and some other types of quantum-circuit \emph{ansatz}. For any particular shift value $s$, there is no other free parameter characterizing the PS estimators. 

In~\cite{Mari:2021estimating}, the \emph{scaled} parameter-shift~(SPS) estimators were introduced. These estimators are essentially scaled versions of the PS estimators, where $\mathrm{SPS}_{\mu,l}f_{\mathrm{Q}}=\lambda\mathrm{PS}_{\mu,l}f_{\mathrm{Q}}$ and $\mathrm{SPS}_{\mu,l}\mathrm{SPS}_{\mu',l'}f_{\mathrm{Q}}=\lambda\mathrm{PS}_{\mu,l}\mathrm{PS}_{\mu',l'}f_{\mathrm{Q}}$ are characterized by an additional prefactor $\lambda$ that ranges from zero to one. For arbitrary shift values of $s$ and prefactor magnitudes, one can similarly arrive at the following TDS accuracy expressions:
\begin{align}
	\MSE{\mathrm{SPS}}{\partial f_\mathrm{Q}}=&\,\overbrace{\dfrac{d\lambda^2}{N_\mathrm{T}(d+1)(\sin s)^2}}^{\displaystyle\mathclap{\text{finite-copy error}}}+\overbrace{\dfrac{d^2(1-\lambda)^2}{2(d+1)(d^2-1)}}^{\displaystyle\text{approx. error}}\,,\nonumber\\
	\MSE{\mathrm{SPS}}{\partial\partial f_\mathrm{Q}}=&\,\dfrac{9d\lambda^2}{8N_\mathrm{T}(d+1)(\sin s)^4}+\dfrac{d^2(1-\lambda)^2}{2(d+1)(d^2-1)}\,,\nonumber\\
	&\,\qquad\qquad{\text{(diagonal Hessian components)}}\nonumber\\
	\MSE{\mathrm{SPS}}{\partial\partial' f_\mathrm{Q}}=&\,\dfrac{d\lambda^2}{N_\mathrm{T}(d+1)(\sin s)^4}+\dfrac{d^4(1-\lambda)^2}{4(d+1)(d^2-1)^2}\,.\nonumber\\
	&\,\qquad\quad\!\!\!{\text{(off-diagonal Hessian components)}}
	\label{eq:SPS_MSE}
\end{align}

It is clear that $s=\pi/2$ will optimize \emph{all} SPS MSEs. We shall see that $\lambda$ can be easily optimized to further enhance estimation accuracies. Notice that the introduction of these prefactors immediately results in the loss of analyticity, since these SPS estimators, for all $\lambda<1$, no longer exactly compute the correct gradient and Hessian components, and therefore carry approximation errors just like the FD and GD estimators. The expressions for general cases are found in Sec.~\ref{app:SPSallcase}. 

\section{Results: Optimally-tuned numerical estimators}
\label{sec:optest}

\subsection{Optimal FD estimators}
\label{subsec:optFD}

From the results in \eqref{eq:FD_MSE}, \eqref{eq:GD_MSE}, \eqref{eq:GD_MSE2} and \eqref{eq:PS_MSE}, the first visual observation is that these formulas are functions of only $N_\mathrm{T}$, the number of qubits $n$ of the circuit, and $\epsilon$. The second observation is that the MSEs are oscillatory functions of $\epsilon$ by virtue of the unitary encoding. The third observation has to do with the choice of $\epsilon$. If one picks very small $\epsilon$ values, then the finite-copy error $\Delta^2_\mathrm{copy}$ dominates as $O(1/\mathrm{poly}~\epsilon)$. If one, instead, picks larger $\epsilon$ values, then the approximation error $\Delta^2_\epsilon$ eventually catches up. The optimal $\epsilon=\epsilon_\mathrm{opt}=\epsilon_\mathrm{opt}(d,N_\mathrm{T})$ minimizes the combination $\mathrm{MSE}=\Delta^2_\mathrm{copy}+\Delta^2_\epsilon$.

Computing $\epsilon_\mathrm{opt}$ through minimizing $\mathrm{MSE}_{\mathrm{FD}}$ over $\epsilon$ requires the solutions of transcendental equations that do not generally admit analytical forms. Numerical optimization methods are therefore in order. Nevertheless, if $[1-\sinc(\epsilon/2)]^2\approx \epsilon^4/576$ and $\{1-[\sinc(\epsilon/2)]^2\}^2\approx \epsilon^4/144$ for small right-hand sides, we may obtain analytical approximations of both $\epsilon_\mathrm{opt}$ and $\mathrm{MSE}^{\mathrm{opt}}_{\mathrm{FD}}$ for optimal~FD gradient and Hessian estimation by approximating $\Delta^2_\epsilon$ to the smallest order $O(\epsilon^4)$. Minimizing the resulting leading-order expansions of all $\mathrm{MSE}_{\mathrm{FD}}$ therefore gives
\begin{align}
	\epsilon_\mathrm{opt}\cong&\,\left(2304\dfrac{d^2-1}{N_{\mathrm{T}}\,d}\right)^{\frac{1}{6}}\quad(\text{$\partial f_\mathrm{Q}$ estimation})\,;\nonumber\\
	\epsilon_\mathrm{opt}\cong&\,\left(5184\dfrac{d^2-1}{N_{\mathrm{T}}\,d}\right)^{\frac{1}{8}}\quad(\text{$\partial\partial f_\mathrm{Q}$ estimation})\,;\nonumber\\
	\epsilon_\mathrm{opt}\cong&\,\left[9216\dfrac{(d^2-1)^2}{N_{\mathrm{T}}\,d^3}\right]^{\frac{1}{8}}\,(\text{$\partial\partial' f_\mathrm{Q}$ estimation})\,.
	\label{eq:optFD}
\end{align}
The corresponding approximately optimal MSEs are
\begin{align}
	\MSE{\mathrm{FD,opt}}{\partial f_\mathrm{Q}}\cong&\,\left(\dfrac{3}{32}\right)^{1/3}\!\!\dfrac{d^{\,4/3}}{(d+1)(d^2-1)^{1/3} N_{\mathrm{T}}^{2/3}}\,,\nonumber\\
	\MSE{\mathrm{FD,opt}}{\partial\partial f_\mathrm{Q}}\cong&\,\dfrac{d^{\,3/2}}{2\,(d+1)(d^2-1)^{1/2} N_{\mathrm{T}}^{1/2}}\,,\nonumber\\
	\MSE{\mathrm{FD,opt}}{\partial\partial' f_\mathrm{Q}}\cong&\,\dfrac{d^{\,5/2}}{3\,(d+1)(d^2-1)N_{\mathrm{T}}^{1/2}}\,.
	\label{eq:optFDMSE}
\end{align}
We note that the formulas in \eqref{eq:optFD} and \eqref{eq:optFDMSE} become accurate so long as $N_\mathrm{T}$ is sufficiently larger than $d$ [that is, $N_\mathrm{T}\gg O(2^n)$]. These formulas can therefore give us a fairly satisfactory description of the estimation errors. 

Perhaps the most striking feature of all $\mathrm{MSE}_{\mathrm{FD,opt}}$s (be it approximated in the $N_\mathrm{T}\gg d$ regime or not) is the fact that they decrease exponentially with the number of qubits $n$. This is desirable since it is shown in Appendices~\ref{app:avgdf} through \ref{app:avgddf} (summarized in Tab.~\ref{tab:avg_res}) that $\overline{\left<(\partial f_\mathrm{Q})^2\right>}$, $\overline{\left<(\partial\partial f_\mathrm{Q})^2\right>}$ and $\overline{\left<(\partial\partial' f_\mathrm{Q})^2\right>}$ are all at most $O(1/d)$, which are manifestations of the so-called barren-plateau phenomenon~\cite{McClean:2018barren,Arrasmith:2021effect,Cerezo:2021cost,Holmes:2022connecting}. Hence, this feature confirms that optimally-tuned FD~estimators are natural for gradient and Hessian estimations. 

\begin{figure*}[t]
	\centering
	\includegraphics[width=2.05\columnwidth]{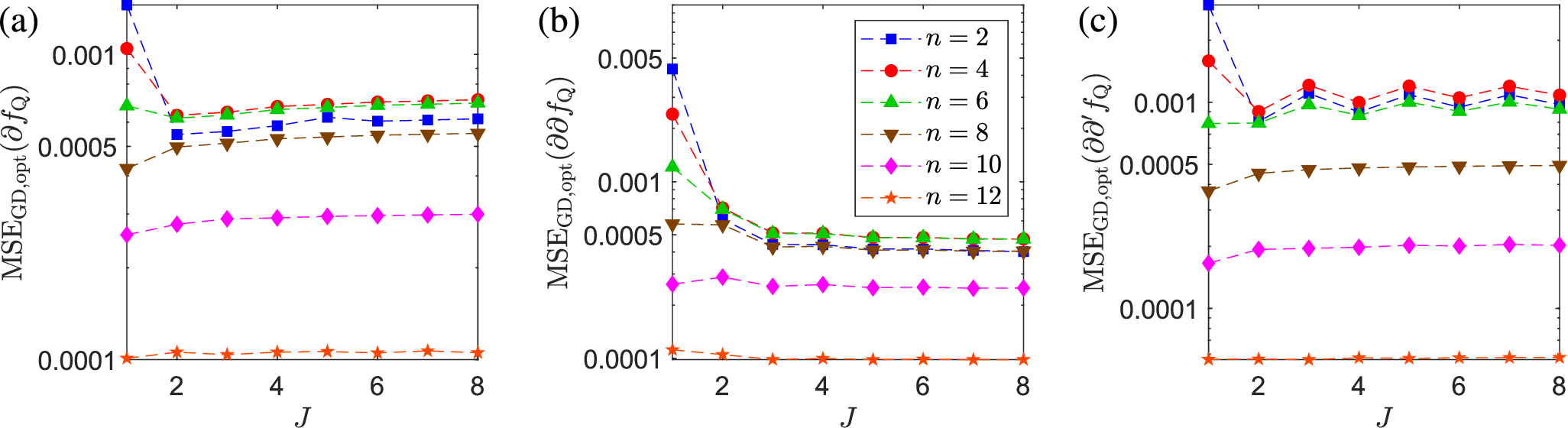}
	\caption{\label{fig:MSEGD}Graphs of $\mathrm{MSE}_{\mathrm{GD,opt}}$ (after analytically minimizing over $\rvec{c}$ in accordance with \eqref{eq:min_c_MSEGD}, followed by numerical minimization over $\epsilon$) for (a)~gradient and (b,c)~Hessian components under the TDS condition defined in Sec.~\ref{subsec:FD}, presented for $1\leq J\leq8$, even number of qubits $n$, and $N_\mathrm{T}=2000$. It turns out that for estimating gradient and off-diagonal Hessian components, $J=2$ appears to be optimal for the regime $n\leq6$, beyond which $J=1$ gives the smallest $\mathrm{MSE}_{\mathrm{GD,opt}}$. For estimating diagonal Hessian components, $J=3$ seems to be the optimal choice, although the differences in estimation errors quickly vanish for large $J$ and $n$ values.} 
\end{figure*}

Ordinarily, things become more difficult to estimate as $n$ increases, so this feature is interestingly counter-intuitive. To understand this behavior, we first note that the average $f_\mathrm{Q}$ landscape rapidly flattens with increasing number of qubits (owing to the barren-plateau phenomenon). For very large $n$, the barren-plateau phenomenon is hence the bottleneck of VQAs. Next, using Tab.~\ref{tab:avg_res}, the optimized FD and PS gradient estimators, for instance, can be shown to have the following average squared magnitudes in the limit of large~$d$: 
\begin{align}
	\text{opt. FD}: &\,\overline{\MEAN{\AVG{}{\widehat{\partial f_\mathrm{Q}}^2}}{}}\rightarrow\left[\sinc(\epsilon_\mathrm{opt}/2)\right]^2\dfrac{1}{2d}+\dfrac{4}{N_\mathrm{T}\epsilon_\mathrm{opt}^2}\,,\label{eq:FDmag}\\
	\text{PS}: &\,\overline{\MEAN{\AVG{}{\widehat{\partial f_\mathrm{Q}}^2}}{}}\rightarrow\dfrac{1}{2d}+\dfrac{1}{N_\mathrm{T}}\rightarrow\dfrac{1}{N_\mathrm{T}}\,.\label{eq:PSmag}
\end{align}	
Since from numerical evidence, $\epsilon_\mathrm{opt}$ grows exponentially for sufficiently large $n$ (refer to the later Sec.~\ref{sec:words} for a concise summary discussion), the average FD estimator squared magnitude drops exponentially in $n$ commensurately with the true components, so that these estimators themselves also approach zero for sufficiently large $n$. As a result, the difference between the estimator and true component converges to zero as $n$ increases. On the other hand, in the large-$d$ limit, \eqref{eq:PSmag} implies that the average PS estimator squared magnitude, and thus the corresponding $\mathrm{MSE}_{\mathrm{PS}}$ is constant for the same $N_\mathrm{T}$.

As a reminder, $\epsilon_\mathrm{opt}$s obtained from minimizing the MSEs in \eqref{eq:FD_MSE} apply when the TDS condition holds (see Sec.~\ref{subsec:FD}). More generally, similar analyses for FD gradient and Hessian estimation that are not under the TDS condition are presented in Appendix~\ref{app:FDallcase}. We shall take the TDS-based formulas in~\eqref{eq:FD_MSE} as representatives useful for obtaining optimal FD~estimators.

We additionally note that the $N_\mathrm{T}^{-2/3}$ and $N_\mathrm{T}^{-1/2}$ scaling behaviors in the first and second equations of \eqref{eq:optFDMSE} were also reported in~\cite{Mari:2021estimating}. The expressions presented there are large-$N$ forms that are not averaged over quantum circuits, and therefore make no reference to average behaviors in $d$ or $n$.

\subsection{Optimal GD estimators}
\label{subsec:optGD}

Optimizing each GD estimator requires the optimization of \emph{both} $\rvec{c}$ and $\epsilon$. For any $\epsilon$, it is easy to carry out the minimization of the MSEs in \eqref{eq:GD_MSE} over normalized $\rvec{c}$, as every GD~MSE takes the form $\mathrm{MSE}_\mathrm{GD}=\rvec{c}^\top\dyadic{M}\,\rvec{c}$. Upon the standard usage of a Lagrange multiplier~(see Appendix~\ref{app:GDallcase}), it is straightforward to arrive at
\begin{equation}
	\min_{\mathclap{\rvec{c}|\,\rvec{c}^\top\onevec=1}}\,\,\mathrm{MSE}_\mathrm{GD}=\left(\onevec^\top\dyadic{M}^{-1}\onevec\right)^{-1}\,,
	\label{eq:min_c_MSEGD}
\end{equation}
where $\rvec{c}_\mathrm{opt}=\dyadic{M}^{-1}\onevec/(\onevec^\top\dyadic{M}^{-1}\onevec)$ and $\dyadic{M}$ is any of the relevant matrices in \eqref{eq:GD_MSE2}. It is clear that the right-hand side of Eq.~\eqref{eq:min_c_MSEGD} reduces to the MSE expressions for FD when $J=1$.

\begin{figure*}[t]
	\centering
	\includegraphics[width=2.05\columnwidth]{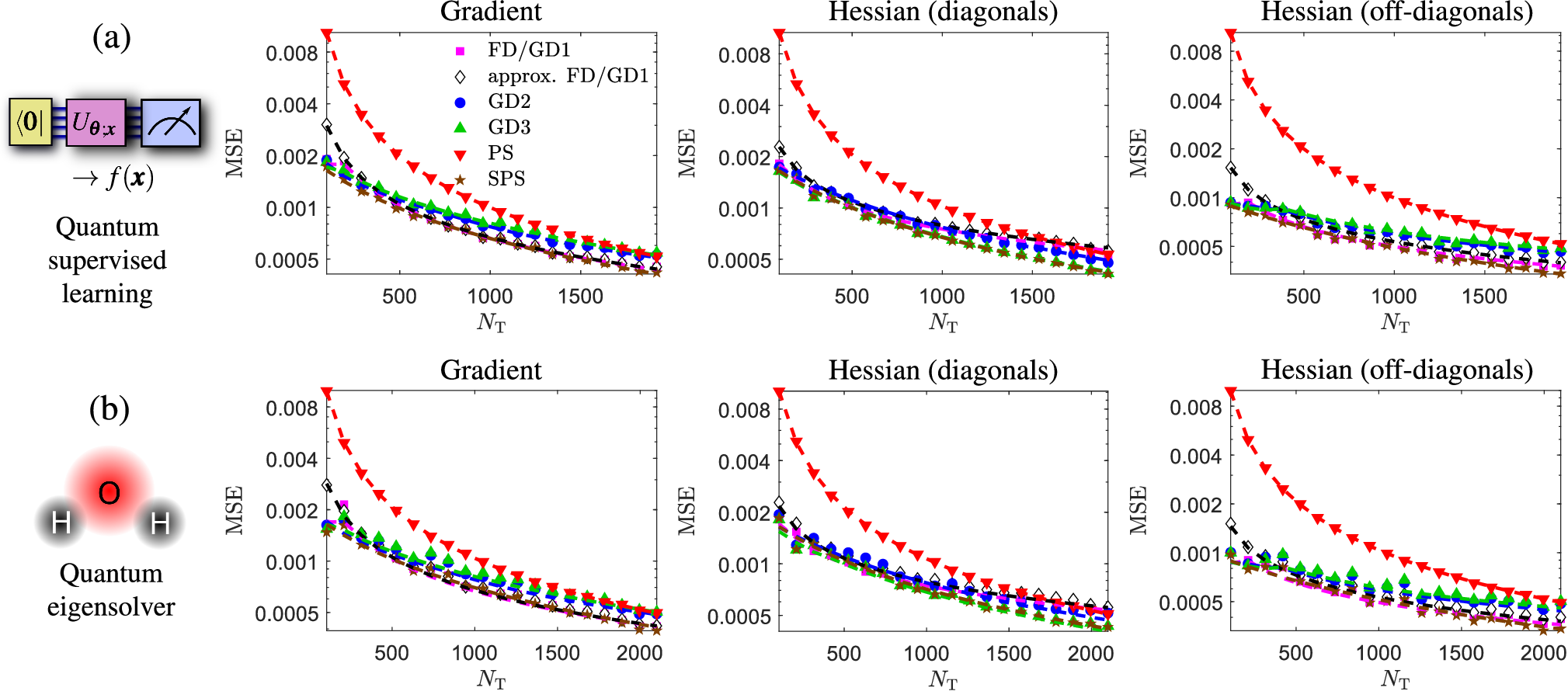}
	\caption{\label{fig:perf_8qbits}Averaged performance plots of gradient- and Hessian-component estimation, generated \emph{via} Monte~Carlo simulations, in the first optimization step of VQAs pertaining to (a)~quantum supervised learning that learns a (multivariate) function $f(\rvec{x})$, and (b)~a quantum eigensolver problem ($V_l=1$ for all $1\leq l\leq L$) that searches for the ground-state energy of a water molecule having restricted excitation levels, with each plot marker averaged over 500 random PEPQCs~($L=5$ and four-repeated units of single-qubit and CNOT gates for each $W_l$) and 500 numerical experiments per PEPQC. All PEPQCs possess the Hilbert-space dimension~$d=2^8$, which are used to estimate $\partial_{50,1}f_\mathrm{Q}$~({\bf Case~II}), $(\partial_{50,1})^2f_\mathrm{Q}$~({\bf Case~II}) and $\partial_{50,1}\partial_{49,2}f_\mathrm{Q}$~({\bf Case~V}). All FD~($\equiv\text{GD1}$ or $J=1$), GD2~($J=2$) and GD3~($J=3$) estimators are optimized according to Sec.~\ref{sec:optest} [``approx.'' refers to the approximated optimization from \eqref{eq:optFD}], where the dashed curves represent the corresponding TDS analytical expressions in \eqref{eq:FD_MSE}, \eqref{eq:GD_MSE}, \eqref{eq:GD_MSE2} and \eqref{eq:PS_MSE}. As the cases (see also Fig.~\ref{fig:cases}) considered here are different from the TDS case ({\bf Case~I}), these TDS curves act as guiding lines, which also show that the actual optimal MSEs do not deviate very far from them. For estimating gradient and off-diagonal Hessian components, both SPS and FD are comparable in performance, whereas it is GD3 that matches with SPS in estimating diagonal Hessian components.} 
\end{figure*}

For $J>1$, there appears to be neither exact nor even approximate analytical forms for $\epsilon_\mathrm{opt}$, although numerical minimization of $\mathrm{MSE}_\mathrm{GD}$s over $\epsilon$ to obtain the optimal $\mathrm{MSE}_\mathrm{GD,opt}$ is efficient so long as $J$ is not too large. Despite the lack of analytical formulas, it is still possible to acquire upper bounds of $\mathrm{MSE}_\mathrm{GD,opt}$ as $N_\mathrm{T}$ and $d$ grows. For this purpose, we first make use of the Cauchy--Schwarz inequality $\vphantom{{W^W}^W}(\onevec^\top\onevec)^2\leq\onevec^\top\dyadic{M}\onevec\onevec^\top\dyadic{M}^{-1}\onevec$ to arrive at the bound
\begin{align}
	\mathrm{MSE}_{\mathrm{GD,opt}}=&\,\min_{\epsilon>0}\quad\!\!\min_{\mathclap{\rvec{c}|\,\rvec{c}^\top\onevec=1}}\mathrm{MSE}_\mathrm{GD}\nonumber\\
	=&\,\min_{\epsilon>0}\left(\onevec^\top\dyadic{M}^{-1}\onevec\right)^{-1}\leq\dfrac{1}{J^2}\min_{\epsilon>0}\onevec^\top\dyadic{M}\onevec\,.
\end{align}
The inequality is saturated when $J=1$. Next, under the TDS condition and $N_\mathrm{T}\gg d$ (or small $\epsilon_\mathrm{opt}$), we can obtain the following explicit upper bounds:
\begin{align}
	&\MSE{\mathrm{GD,opt}}{\partial f_\mathrm{Q}}\lesssim\dfrac{1}{4}\left[\dfrac{d^4(J+1)^2(2J+1)^2}{6\,N_\mathrm{T}^2J^2(d+1)^3(d^2-1)}\right]^{1/3}\!\!\HARM{J,2}^{2/3}\,,\nonumber\\
	&\MSE{\mathrm{GD,opt}}{\partial\partial f_\mathrm{Q}}\lesssim\frac{J+1}{36\,J}\left(\dfrac{2dJ+d}{d+1}\right)^{\!\!3/2}\!\!\sqrt{\dfrac{2\HARM{J,2}^2\!\!+\!\!\HARM{J,4}}{N_\mathrm{T}(d-1)}}\,,\nonumber\\
	&\MSE{\mathrm{GD,opt}}{\partial\partial' f_\mathrm{Q}}\lesssim\dfrac{(J+1)(2J+1)}{18(d+1)(d^2-1)}\left(\dfrac{d^5\HARM{J,4}}{N_\mathrm{T}J}\right)^{\!\!1/2}\,,
	\label{eq:optGD_UB}
\end{align}
where $\HARM{j,k}=\sum^j_{m=1}m^{-k}$ is the generalized harmonic number. It is obvious that all right-hand sides reduce to the expressions in \eqref{eq:FD_MSE} when $J=1$. Based on these crude upper bounds, one can already deduce the fundamental trend: $\mathrm{MSE}_{\mathrm{GD,opt}}$ decreases with increasing $d$ and $N_\mathrm{T}$, which comes as no surprise on hindsight since the GD~strategy is a generalization of the FD~strategy, and naturally preserves the exponential-decay-in-$n$ characteristic.

For any $d$ and $N_\mathrm{T}$, resorting to numerical optimization for deciding on the optimal value of $J$ is inevitable. For a prechosen $N_\mathrm{T}$, number of qubits~$n$ employed by the quantum circuit and the type of components (gradient or Hessian) to be estimated, the optimal value of $J$ is the one that minimizes the numerical minimum $\mathrm{MSE}_{\mathrm{GD,opt}}$. As examples, Fig.~\ref{fig:MSEGD} illustrates the graphs of the respective $\mathrm{MSE}_\mathrm{GD,opt}$s for $N_\mathrm{T}=2000$ using various $n$ values. In these cases, there exist optimal values of $J$ below a certain critical $n$. Beyond this critical value, $J=1$ is sufficient as larger $J$ values generally do not significantly vary $\mathrm{MSE}_\mathrm{GD,opt}$.

\subsection{Optimal SPS estimators}

The minimization of \eqref{eq:SPS_MSE} with respect to $\lambda$, unlike the FD and GD strategies, can be carried out \emph{exactly} for \emph{any} $d$ and $N_\mathrm{T}$ since this simply amounts to the minimization of quadratic functions in $\lambda$ on the right-hand sides. Hence, one arrives at the optimal scaling prefactors
\begin{align}
	\lambda_{\mathrm{opt}}=&\,\dfrac{dN_\mathrm{T}}{2d^2 + dN_\mathrm{T}-2}\quad\!\qquad(\text{$\partial f_\mathrm{Q}$ estimation})\,,\nonumber\\
	\lambda_{\mathrm{opt}}=&\,\dfrac{4dN_\mathrm{T}}{9d^2 + 4dN_\mathrm{T}-9}\,\,\qquad(\text{$\partial\partial f_\mathrm{Q}$ estimation})\,,\nonumber\\
	\lambda_{\mathrm{opt}}=&\,\dfrac{d^3N_\mathrm{T}}{4(d^2-1)^2 + d^3N_\mathrm{T}}\,\quad(\text{$\partial\partial' f_\mathrm{Q}$ estimation})\,,
	\label{eq:SPS_lbd_opt}
\end{align}
and realizes that $\lambda_{\mathrm{opt}}\rightarrow1$ as $d\ll N_\mathrm{T}\rightarrow\infty$, and $\lambda_{\mathrm{opt}}\rightarrow0$ as $N_\mathrm{T}\ll d\rightarrow\infty$. The former limit for any fixed~$d$ is obvious, since in the large-data limit, scaling prefactors are not really necessary and all SPS estimation performances approach to those of PS. On the other hand, for a fixed $N_\mathrm{T}$, sampling from a circuit that has a large number of qubits would pay off with very small prefactors.

Consequently, the corresponding optimized MSE expressions for these optimized SPS estimators are given by
\begin{align}
	\MSE{\mathrm{SPS,opt}}{\partial f_\mathrm{Q}}=&\,\dfrac{d^2}{(d+1) (2d^2 + dN_\mathrm{T}-1)}\,,\nonumber\\
	\MSE{\mathrm{SPS,opt}}{\partial\partial f_\mathrm{Q}}=&\,\dfrac{9d^2}{2 (d+1) (9d^2 + 4dN_\mathrm{T}-9)}\,,\nonumber\\
	&\,\qquad\,{\text{(diagonal Hessian components)}}\nonumber\\
	\MSE{\mathrm{SPS,opt}}{\partial\partial' f_\mathrm{Q}}=&\,\dfrac{d^4}{(d+1) [4(d^2-1)^2 + d^3N_\mathrm{T}]}\,.\nonumber\\
	&\,\quad\!\!{\text{(off-diagonal Hessian components)}}
	\label{eq:SPS_MSE_opt}
\end{align}
For a fixed $d$, these optimized SPS MSEs converge to the PS MSEs in \eqref{eq:PS_MSE}, as they should.

\section{Performance}
\label{sec:perf}

\subsection{Applications in quantum supervised learning and quantum eigensolver problems}

We shall first demonstrate the performance of optimal FD and GD estimators in typical applications of VQAs. As a benchmark, we compare the estimation errors of these optimal estimators with those from PS estimators, using the MSE as figure of merit. The first important example of a VQA is quantum machine learning, where a PEPQC that defines the quantum model $f_\mathrm{Q}(\rvec{x})=\opinner{\rvec{0}}{U_{\rvec{\theta};\rvec{x}}^\dag O U_{\rvec{\theta};\rvec{x}}}{\rvec{0}}$ is trained to learn or express a general multivariate function $f(\rvec{x})$ [$|f(\rvec{x})|\leq1$ for all $\rvec{x}$ with no loss of generality] by minimizing an appropriate cost function. In this case, it is sufficient to assign the observable circuit $O$ as the multiqubit Pauli operator $Y\otimes1^{n-1}$. 

The second widely-studied example concerns quantum-eigensolver tasks that search for minimum eigenvalues of operators. Here, the observable $O$ is one such operator of interest that is typically Hermitian (for example, a Hamilton operator describing the dynamics of an electronic system in a molecule), and hence, decomposable as a linear combination of multiqubit Pauli operators $O_k$. For this second application, we shall consider a simplified (trace-subtracted) Hamilton operator $O$ that describes the electrons in a water molecule with restricted excitation levels. Using the Jordan--Wigner transformation, one may write $O$ as a weighted sum of 96 eight-qubit Pauli operators~(see Appendix~\ref{app:water}). 

Figure~\ref{fig:perf_8qbits} showcases the estimation-error performances of gradient and Hessian estimation in some of the physical cases listed in Fig.~\ref{fig:cases} for the two aforementioned examples of VQA applications. As a benchmark, the results indeed confirm that optimal FD and GD strategies outperform the PS strategy for $N_\mathrm{T}$ below certain critical values that would depend on $d$ and the types of components. We also remark that although the analytical curves in Fig.~\ref{fig:perf_8qbits} are strictly meant for components under \textbf{Case~I}~in Fig.~\ref{fig:cases} that is equivalent to the TDS condition, we observe, through these and other numerical evidence not shown here, that the analytical results in \eqref{eq:FD_MSE} and \eqref{eq:GD_MSE} supplying those curves can also approximate the estimation errors for other cases well. Another important sanity verification from the figure is that the optimal FD approximators defined by the respective $\epsilon_\mathrm{opt}$s prescribed in \eqref{eq:optFD}, along with their MSEs in \eqref{eq:optFDMSE}, quickly converge to the exact optimal curves with increasing $N_\mathrm{T}$.

The estimation accuracy greatly improves when analyticity is forsaken in SPS, where the optimization of the respective scaling prefactors with respect to the averaged MSEs offer comparable estimation performances with the optimal FD and GD strategies. As $N_\mathrm{T}$ increases, SPS eventually becomes the most efficient strategy.

\subsection{Benefits of optimized numerical estimators for scalable NISQ devices}

Relative to the standard PS strategy, the scaled version, SPS, is statistically more efficient in estimating gradient and Hessian components. This is immediately clear from either a direct comparison of \eqref{eq:SPS_MSE_opt} with \eqref{eq:PS_MSE} under the TDS condition, or the simple arguments in Sec.~\ref{app:SPSallcase} for all other cases: for \emph{any} $d\geq2$ and $N_\mathrm{T}>0$, $\MSE{\mathrm{SPS}}{\,\cdot\,}<\MSE{\mathrm{PS}}{\,\cdot\,}$.

That the optimized FD and GD strategies could give smaller estimation errors than PS for a significantly large regime of $N_\mathrm{T}$ can be understood by noting that the PS estimators defined in Eqs.~\eqref{eq:PS_est3} and \eqref{eq:PS2_est3} correspond to an effective $\epsilon\cong\pi$ that is generally not optimal for minimizing the estimation error---the \emph{combined} errors $\Delta^2_\mathrm{copy}$ and $\Delta^2_\epsilon$. When $N_\mathrm{T}$ is greater than a certain critical value $N_*$ (defined as the value of $N_\mathrm{T}$ for which $\mathrm{MSE}_\mathrm{FD/GD,opt}=\mathrm{MSE}_\mathrm{PS}$), the contribution of $\Delta^2_\mathrm{copy}$ is small enough to be dominated by the nonzero-$\epsilon$ error $\Delta^2_\epsilon$, such that the advantage of an $\epsilon$-error-free PS estimator manifests itself with a smaller MSE relative to those of the FD or GD approximators.

\begin{figure*}[t]
	\centering
	\includegraphics[width=2\columnwidth]{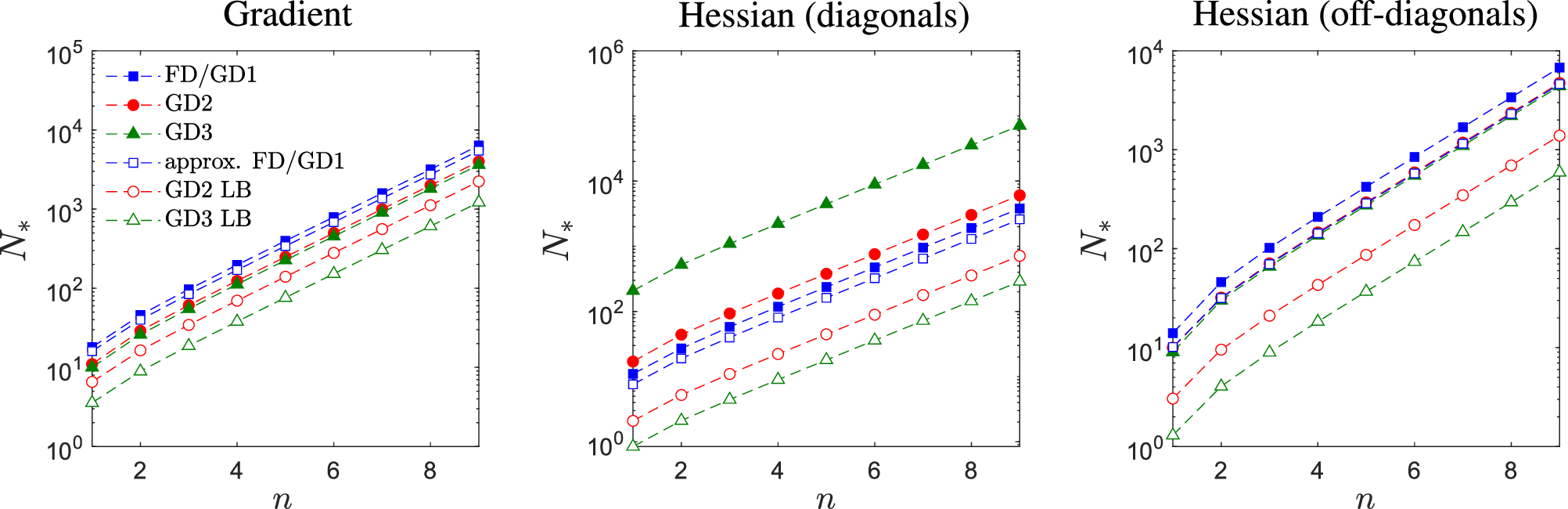}
	\caption{\label{fig:N_star}Exponential increase of $N_*$ in gradient and Hessian estimation schemes under the TDS condition. The $N_*$ values corresponding to the approximated FD~strategy~[\eqref{eq:N_star_FD}], and those given by the cruder lower bounds of both the GD2~[$J=2$, GD2~LB] and GD3~[$J=3$, GD3~LB] strategies in \eqref{eq:N_star_GD}, all serve as lower bounds to the actual $N_*$ values computed respectively using the correct $\epsilon_\mathrm{opt}$s obtained from numerical optimization of MSEs in \eqref{eq:FD_MSE} and \eqref{eq:GD_MSE} over $\epsilon$. For each estimation scheme, a larger $N_*$ is indicative of a greater sampling efficiency over the PS strategy.} 
\end{figure*}

To further support the usefulness of optimal FD and GD schemes, we answer the important question: ``How does $N_*$ scale with the number of qubits $n$?''. Basic intuition suggests that since the FD and GD estimation errors decreases with $n$ according to Sec.~\ref{subsec:optFD}, while the PS ones \emph{do not} [recall~\eqref{eq:PS_MSE}], the critical value $N_\mathrm{T}=N_*$ required for the PS strategy to start outperforming the former schemes would also grow with $n$. Indeed, based on the approximately minimized MSE's in \eqref{eq:optFDMSE} for the FD~strategy under the TDS condition and the regime $N_*\gg d$, we find that
\begin{align}
	N_*\cong&\,\dfrac{32(d^2-1)}{3d}\quad(\text{$\partial f_\mathrm{Q}$ estimation})\,;\nonumber\\
	N_*\cong&\,\dfrac{81(d^2-1)}{16d}\quad(\text{$\partial\partial f_\mathrm{Q}$ estimation})\,;\nonumber\\
	N_*\cong&\,\dfrac{9(d^2-1)^2}{d^3}\quad(\text{$\partial\partial' f_\mathrm{Q}$ estimation})\,.
	\label{eq:N_star_FD}
\end{align}
For the GD~strategy under the TDS condition, the upper bounds derived and stated in \eqref{eq:optGD_UB} conveniently permit us to write down \emph{approximate and loose lower bounds} of $N_*$:
\begin{align}
	N_*\gtrsim&\,\dfrac{384J^2(d^2-1)}{d\,(J+1)^2(2J+1)^2\,\HARM{J,2}^2}\!\!\qquad\qquad(\text{$\partial f_\mathrm{Q}$ estimation})\,;\nonumber\\
	N_*\gtrsim&\,\dfrac{6561J^2(d^2-1)}{4\,d\,(J+1)^2(2J+1)^3(2\HARM{J,2}^2+\HARM{J,4})}\quad(\text{$\partial\partial f_\mathrm{Q}$ est.})\,;\nonumber\\
	N_*\gtrsim&\,\dfrac{6561J\,(d^2-1)^2}{16\,d^3\,(J+1)^2(2J+1)^2\,\HARM{J,4}}\,\quad(\text{$\partial\partial' f_\mathrm{Q}$ estimation})\,.
	\label{eq:N_star_GD}
\end{align}

While Eqs.~\eqref{eq:N_star_FD} and \eqref{eq:N_star_GD} strictly hold only when $N_*\gg d$, they analytically show, at least in this regime, that $N_*\gtrsim O(2^n)$. The plots in Fig.~\ref{fig:N_star} clearly shows an exponential increase in $N_*$ with respect to $n$ regardless of whether $\epsilon_\mathrm{opt}$ is found with numerical MSE optimization or large $N_\mathrm{T}$ approximation. In particular, for the optimized FD strategy, the differences between the exact numerically obtained $N_*$s and those from \eqref{eq:N_star_FD} are very small. In general, $N_*$ is a useful measure for the sampling efficiency of a particular optimized scheme in question. A larger $N_*$ implies that the optimized numerical scheme gives a smaller estimation error for a larger range of $1\leq N_\mathrm{T}\leq N_*$ in contrast to the analytical PS strategy. A scheme that exhibits an exponentially growing $N_*$ with respect to $n$ is therefore a much more statistically favorable one over~PS for scalable VQAs. Figure~\ref{fig:N_star} illustrates the exponential growth in $N_*$ with respect to $n$ for estimations performed under the TDS condition. More general arguments in Appendices~\ref{app:QCavg} and~\ref{app:smp_err}, which are applicable to \emph{all cases} in Fig.~\ref{fig:cases}, technically guarantee an exponentially growing~$N_*$ with $n$ for \emph{all}~FD and GD~estimation schemes. With this, optimally-tuned FD and GD strategies may be regarded as prime candidates for scalable VQAs, especially on NISQ platforms where estimating large-$n$ circuit-model expectation values with large numbers of sampling copies is practically infeasible.

\section{Important remarks and potential pitfall}
\label{sec:words}

The results in this article show that numerical estimators possessing free parameters~($\epsilon$ for FD and GD and $\lambda$ for SPS) can be optimized to yield more accurate gradient and Hessian estimation than analytical estimators~(PS) that do not possess such a freedom. The optimization refers to the minimization of the relevant circuit-averaged MSE---an estimation-error quantifier for the circuit function, gradient and Hessian---of a given circuit \emph{ansatz} and sampling-copy number $N_\mathrm{T}$ with respect to the free parameter. We recall that the average is performed over not just the ``click'' data per training circuit, but also over all possible training-circuit parameters according to the \emph{ansatz} structure. 

For optimized FD and GD schemes, we reiterate that $\epsilon_\mathrm{opt}$ is consequently dependent on the circuit \emph{ansatz} and $N_\mathrm{T}$. For the PEPQC \emph{ansatz} that leads to two-design approximating modules considered in this work, the resulting MSE expressions, as shown for instance in Eq.~\eqref{eq:FD_MSE} and \eqref{eq:GD_MSE}, are nonlinear functions of $\epsilon$, so that numerical methods for their minimization are the only resort. Otherwise, analytical approximations of $\epsilon_\mathrm{opt}$ such as those in~\eqref{eq:optFD} for the FD strategy may be employed when $N_\mathrm{T}\gg d$.

\begin{figure*}[t]
	\centering
	\includegraphics[width=2.05\columnwidth]{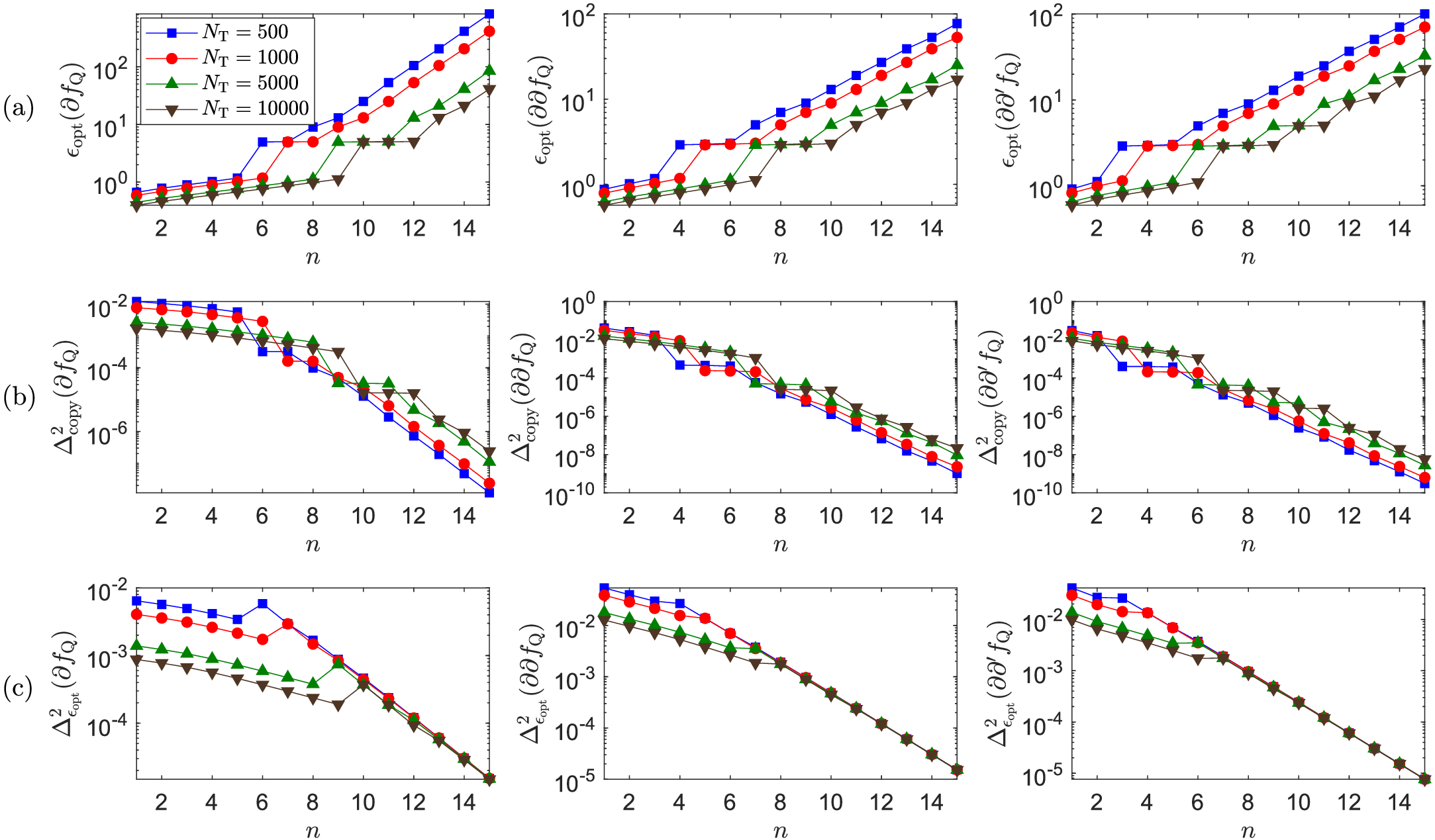}
	\caption{\label{fig:epsopt_cpy_app}Graphs of (a)~$\epsilon_\mathrm{opt}$, (b)~finite-copy error $\Delta^2_\mathrm{copy}$ and (c)~nonzero-$\epsilon_\mathrm{opt}$ approximation error $\Delta^2_{\epsilon_\mathrm{opt}}$ for all three types of components estimated with the FD strategy under various qubit numbers $n$ and $N_\mathrm{T}$. All $\epsilon_\mathrm{opt}$s are obtained by numerically minimizing the respective TDS MSE expressions in \eqref{eq:FD_MSE} with \emph{no} approximations. Numerical evidence shows that $\epsilon_\mathrm{opt}$ ultimately grows exponentially in $n$. At the same time, we see that both $\Delta^2_{\mathrm{copy}}$ and $\Delta^2_{\epsilon_\mathrm{opt}}$ also asymptotically drop exponentially with $n$. Note that the exponential drop in $\Delta^2_{\epsilon_\mathrm{opt}}$ is solely due to $n$. As a special case, when $n$ is small so that $N_\mathrm{T}\gg d$, $\epsilon_\mathrm{opt}$ is clearly small. Otherwise, the magnitude of $\epsilon_\mathrm{opt}$ can be large by virtue of the behaviors of $\Delta^2_{\mathrm{copy}}$ and $\Delta^2_{\epsilon_\mathrm{opt}}$.} 
\end{figure*}

Moreover, rather unintuitively at first glance, we find that $\epsilon_\mathrm{opt}$ is typically not small, regardless of whether numerical optimization or analytical approximations are invoked. In fact, as $d$ increases, numerical experience shows that $\epsilon_\mathrm{opt}$ grows roughly exponentially with $n$. One can already witness this behavior approximately from~\eqref{eq:optFD}. This is, apparently, at odds with the usual narrative that $\epsilon_\mathrm{opt}$ should be as small as possible, preferably $\epsilon_\mathrm{opt}\rightarrow0$, in order for the approximation of gradient and Hessian components to be as ``accurate'' as possible. While such a narrative is surely correct when no sampling is required to estimate the components, in which case one should just use the PS scheme and not even be bothered with FD, GD or any other numerical scheme, matters greatly differ when sampling is required, as in the case of VQAs. For FD and GD, it is obvious (see also the start of Sec.~\ref{subsec:optFD}) that $\epsilon_\mathrm{opt}$ is neither minuscule nor astronomical, but somewhere in between. However, for a fixed $N_\mathrm{T}$, as $d$ or $n$ increases, $\epsilon_\mathrm{opt}$ tends to larger values because the nonzero-$\epsilon_\mathrm{opt}$ approximation error $\Delta_{\epsilon_\mathrm{opt}}^2$ approaches zero while the finite-copy error $\Delta_\mathrm{copy}^2$ is asymptotically constant. The value of $\epsilon_\mathrm{opt}$ becomes tiny \emph{only when} $N_\mathrm{T}$ is astronomical, in which case the usual narrative applies. Figure~\ref{fig:epsopt_cpy_app} visually demonstrates all these remarks for the FD strategy as an example. 

If one, for instance, inspects the magnitudes of the respective gradient estimators for FD and PS, one arrives at the large-$d$ limit formulas in \eqref{eq:FDmag} and \eqref{eq:PSmag} for any fixed $N_\mathrm{T}$. On the other hand, the barren-plateau phenomenon for the PEPQC \emph{ansatz} implies that $\overline{\MEAN{\left(\partial f_\mathrm{Q}\right)^2}{}}$ drops exponentially in $n$ (see Tab.~\ref{tab:avg_res}). Therefore, given a fixed $N_\mathrm{T}$, the presence of $\epsilon_\mathrm{opt}=\epsilon_\mathrm{opt}(d,N_{\mathrm{T}})$ in step-size dependent strategies such as FD and GD can introduce a faster diminishing average gradient-estimator magnitude that is more compatible with the barren-plateau phenomenon, especially when $d$ is large. More specifically, these strategies do so \emph{not} by choosing some \emph{ad hoc} $\epsilon_\mathrm{opt}$ that is large, of course, but by minimizing either the TDS MSE, or MSE upper bounds in other non-TDS cases, provided that the circuit \emph{ansatz} is known beforehand---the PEPQC \emph{ansatz} in our case. The PS strategy lacks this additional parameter degree of freedom, and therefore depends only on a sufficiently large $N_\mathrm{T}$ to surpass the performances of FD and GD on average.

We forewarn that while optimized numerical estimators can boost estimation accuracies, this \emph{does not} necessarily mean that these estimators will improve circuit trainability under the influence of the barren-plateau phenomenon. These are clearly two separate problems, the latter of which is \emph{not} addressed by this article. The analysis of the MSE throughout the manuscript reveals the statistical quality in estimating gradients and Hessians. The appearance of a free parameter that characterizes a numerical estimator permits further estimation-accuracy improvement with proper optimization of this parameter. For an extremely large number of qubits $n$, the landscape of $f_\mathrm{Q}(\rvec{\theta};\rvec{x})$ from a randomized two-design circuit, for instance, is almost flat, so that the problem of \emph{distinguishing} function magnitudes in different $\rvec{\theta}$ search directions still persists, even with accurate statistical estimations. 

An absurd hypothetical scenario would be $n\rightarrow\infty$, where the initialized $\MEAN{f_\mathrm{Q}^2}{}\cong0\cong f_\mathrm{Q}$, while the optimized estimation error of any numerical estimator (FD, GD or SPS) is nearly zero [notice that from~\eqref{eq:PS_MSE}, analytical PS estimators still give $O(1/N_\mathrm{T})$ estimation errors that are not necessarily small in this scenario]. Such an infinitely-large quantum circuit is not trainable, so low estimation errors surely \emph{does not} translate to better trainability. To see why, we note that trainability pertains to function-minimization efficiency, and so one should be strict about picking the right descent direction in every function-value update. For a very large $n$, such that the true gradient has a tiny magnitude, a very small gradient-estimation MSE could still lead to many wrong update directions with even slight statistical fluctuations, so that cost-function minimization can still be very slow on average. Thus, one should, instead, find ways to, \emph{as an example}, reduce
\begin{equation}
	\mathcal{D}_{\rvec{\theta}_0}=\MEAN{\dfrac{\max\left\{\VAR{}{\widehat{f}_\mathrm{Q}(\rvec{\theta}\pm\rvec{\theta}_0;\rvec{x})}\right\}}{\left|f_\mathrm{Q}(\rvec{\theta}+\rvec{\theta}_0;\rvec{x})-f_\mathrm{Q}(\rvec{\theta}-\rvec{\theta}_0;\rvec{x})\right|^2}}{}
\end{equation}
for some chosen displacement $\rvec{\theta}_0$ when speaking of trainability. This quantifies the worst-case average relative spread~(variance) of $\widehat{f}_\mathrm{Q}(\rvec{\theta}\pm\rvec{\theta}_0;\rvec{x})$, which if too large, results in the failure of distinguishing between $f_\mathrm{Q}(\rvec{\theta}+\rvec{\theta}_0;\rvec{x})$ and $f_\mathrm{Q}(\rvec{\theta}-\rvec{\theta}_0;\rvec{x})$ in the direction set by $\rvec{\theta}_0$ through sampling. For quantum circuits of large~$d$ that exhibit two-design properties, the numerator approaches $O(1/N)$ for $N$ sampling copies, and the denominator is at most $O(1/\poly~d)$, so that $\mathcal{D}_{\rvec{\theta}_0}\gtrsim O((\poly~d)/N)$. For large circuits, $N$ must thus at least be exponentially large in $n$ for trainability~\cite{Zhang:2020toward}.

Efforts in ameliorating the effects of barren plateaus would therefore require deeper understanding in both the circuit \emph{ansatz}~\cite{Zhang:2020toward,Zhao:2021analyzing}, and $\rvec{\theta}$-initialization and optimization strategies~\cite{Haug:2021optimal}, for instance. Improving model trainability in the presence of barren plateaus is a pertinent task without a doubt, but is beyond the scope of this article. With that said, it cannot be overemphasized that \emph{both} accurate estimation and trainability are equally important in quantum computation especially in the NISQ era where every bit of noise counts. The \emph{methodology} for optimizing estimators may be extended to other circuit \emph{ans{\"a}tze} and circuit-parameter initialization procedures beyond the two-design approximating \emph{ans{\"a}tze} and randomized initialization considered in this article.

\section{Conclusion}

Executing variational algorithms on modern NISQ devices typically necessitate the computation of circuit-function gradient and Hessian components through direct variational-circuit-function sampling. A thorough understanding of the inherent estimation errors is vital to ensure the reliability of NISQ computation. In this work, we provide detailed analyses on the estimation errors for various gradient and Hessian computation methods that are relevant not only to gradient and Hessian-assisted optimization approaches, but also nongradient-based routines which require the estimation of circuit-function differences.

Armed with these fundamental results that apply to very general variational quantum-computation settings, we propose optimally-tuned gradient and Hessian numerical estimators that offer significantly-reduced average estimation errors on any NISQ device that can only supply a finite number of sampling copies within a given operation time duration. These optimized numerical estimators work especially well in improving the gradient and Hessian computation accuracies during the initial stages of cost-function optimization, where training parameters are first randomly initialized before the optimization procedure such that all polynomially-deep training modules possessing a ``hardware-efficient'' {\it ansatz} behave closely as quantum two-designs. The simulation results suggest that such optimally-tuned estimators are still extremely advantageous in estimation-error minimization for training modules that are shallow.

Moreover, these numerical estimators are compatible with the barren-plateau phenomenon; that is, given a fixed number of sampling copies, the average estimation errors based on these optimized estimators scale with the corresponding root-mean-squares of circuit-function gradient and Hessian components, both of which drop exponentially with the number of qubits employed. This desirable feature prevents all gradient and Hessian computation from turning into random guesses for a fixed number of sampling copies as the quantum-circuit size increases. 

For the \emph{same} number of sampling copies, this is in contrast to the analytical unscaled parameter-shift rule, which estimates gradients and Hessians with errors that are asymptotically independent of the circuit-qubit number. We showed that this consequently requires an exponentially-increasing number of sampling copies with the qubit number in order for the analytical estimators to overtake the corresponding optimally-tuned ones in sampling performance. Hence, while the absence of approximation errors with this analytical rule is a commonly sought-after characteristic, optimally-tuned numerical estimators (including the scaled parameter-shift estimators) still present a much more feasible estimation strategy on practical NISQ devices where sampling-copy numbers are finite.

An obvious next step to take towards practical applications would be a performance analysis of known sampling strategies in the presence of realistic noisy environments that perforate typical quantum-computing architectures, such as photon loss and depolarization. Knowledge about how (potentially biased) noise models influence sampling computation is pertinent for proposing possibly noise-model-agnostic optimized strategies to improve estimation qualities. Another interesting area of discussion begins with recognizing the error-mitigating effects in using the unscaled parameter-shift rule, the reason of which is due to a large fixed step size to define gradient and Hessian components, which could overlook slight noise perturbations in the components and render this analytical rule robust against noise. In contrast, conventional wisdom often suggests that finite-difference methods employ much smaller step sizes leading to estimators that are relatively less robust to noise. However, when the knowledge about the circuit \emph{ansatz} is accounted for, the resulting optimized finite-difference strategies correspond to step-size magnitudes that could be comparable with those of the unscaled parameter-shift rule. Hence, the study of possible error-mitigative power for optimized numerical schemes shall certainly be a part of the immediate future agenda.

\begin{acknowledgements}
	The author thanks S.~Shin for fruitful discussions. This work is supported by the National Research Foundation of Korea (NRF) grants funded by the Korea government (Grant nos.~NRF-2020R1A2C1008609, NRF-2020K2A9A1A06102946, NRF-2019R1A6A1A10073437 and NRF-2022M3E4A1076099) \emph{via} the Institute of Applied Physics at Seoul National University.
\end{acknowledgements}

\appendix

\section{Basic properties of PEPQCs}
\label{app:basic_props}

\subsection{Haar-measure integration for quantum two-designs}

An $n$-qubit serial-circuit model that contains trainable unitary modules $W_l$ that each has a $O(\poly(n,2))$ circuit depth, so that a randomized $W_l$ for a broad class of circuit {\it ans{\"a}tze} (including circuits consisting of single-qubit and CNOT gates) may be approximated as a two-design~\cite{harrow_random_2009}. PEPQCs form a subclass of such two-design approximating circuits.

In view of this, the following integration result
\begin{align}
	&\,\int(\D U)_\mathrm{Haar} U^*_{j'_1k'_1}U^*_{j'_2k'_2}U_{j_1k_1}U_{j_2k_2}\nonumber\\
	=&\,\dfrac{\delta_{j_1,j'_1}\delta_{j_2,j'_2}\delta_{k_1,k'_1}\delta_{k_2,k'_2}+\delta_{j_1,j'_2}\delta_{j_2,j'_1}\delta_{k_1,k'_2}\delta_{k_2,k'_1}}{d^2-1}\nonumber\\
	&\,-\dfrac{\delta_{j_1,j'_1}\delta_{j_2,j'_2}\delta_{k_1,k'_2}\delta_{k_2,k'_1}+\delta_{j_1,j'_2}\delta_{j_2,j'_1}\delta_{k_1,k'_1}\delta_{k_2,k'_2}}{d(d^2-1)}
	\label{eq:Weingarten2}
\end{align}
in terms of the computational matrix elements $U_{jk}=\opinner{j}{U}{k}$ of a $d$-dimensional random unitary operator $U$ distributed according to the Haar measure $(\D U)_\mathrm{Haar}$ and the basic identity
\begin{equation}
	\Haar{U\,O\,U^\dag}=\int(\D U)_\mathrm{Haar}\,U\,O\,U^\dag=\frac{1}{d}\tr{O}\,,
\end{equation}
are relevant~\cite{Puchala_Z._Symbolic_2017}. By tracking all indices, it is possible to derive another useful integral identity
\begin{align}
	&\,\Haar{U^{\otimes2}\,O\, U^{\dag\,\otimes2}}=\int(\D U)_\mathrm{Haar}\,U^{\otimes2}\,O\, U^{\dag\,\otimes2}\nonumber\\
	=&\,\left[\dfrac{\tr{O}}{d^2-1}-\dfrac{\tr{O\tau}}{d(d^2-1)}\right]1+\left[\dfrac{\tr{O\tau}}{d^2-1}-\dfrac{\tr{O}}{d(d^2-1)}\right]\tau\,,
	\label{eq:HaarV2}
\end{align}
where $\tau$ is the swap operator that carries the simple trace property $\tr{O_1\otimes O_2\,\tau}=\tr{O_1O_2}=\tr{U^{\otimes2}\,O_1\otimes O_2\,U^{\dag\otimes2}\,\tau}$ for any two observables $O_1$ and $O_2$, and unitary operator $U$. If one observable is traceless,
\begin{equation}
	\Haar{U^{\otimes2}\,O_1\otimes O_2\, U^{\dag\,\otimes2}}=\dfrac{d\,\tau-1}{d(d^2-1)}\,\tr{O_1O_2}\,.
	\label{eq:pauliV2}
\end{equation} 

\subsection{Training-parameter translation in $f_{\mathrm{Q},k}$}
\label{subapp:translation}

If we denote $A=\prod^{l+1}_{l'=L}V_{l'}W_{l'}$ and $B=\prod^{1}_{l'=l-1}V_{l'}W_{l'}$, then
\begin{align}
\partial_{\mu,l}f_{\mathrm{Q},k}=&\,\dfrac{\I}{2}\opinner{\rvec{0}}{B^\dag W^{(2)\dag}_l\sigma_{\mu l} W^{(1)\dag}_lV_l^\dag A^\dag O_k AV_lW_lB}{\rvec{0}}\nonumber\\
&\,+\mathrm{c.c.}\,,\nonumber\\
\left(\partial_{\mu,l}\right)^2f_{\mathrm{Q},k}=&\,\dfrac{1}{2}\bra{\rvec{0}}B^\dag W^{(2)\dag}_l\sigma_{\mu l} W^{(1)\dag}_lV_l^\dag A^\dag O_k AV_l\nonumber\\
&\,\quad\times W^{(1)}_l\sigma_{\mu l} W^{(2)}_lB\ket{\rvec{0}}-\dfrac{1}{2}f_{\mathrm{Q},k}\,,
\label{eq:df,d2f}
\end{align}
where the argument $\rvec{x}$ is hereby unstated for notational simplicity unless otherwise necessary. From the unique property $\sigma^2_{\mu l}=1$ of (multiqubit) Pauli operators employed in PEPQCs, all higher-order derivatives are simply $\partial_{\mu,l}f_{\mathrm{Q},k}$ and $(\partial_{\mu,l})^2f_{\mathrm{Q},k}$ multiplied by simple phase factors:
\begin{align}
	\left(\partial_{\mu,l}\right)^{2k+1}f_{\mathrm{Q},k}=&\,(-1)^k\partial_{\mu,l}f_{\mathrm{Q},k}\,,\nonumber\\
	\left(\partial_{\mu,l}\right)^{2k}f_{\mathrm{Q},k}=&\,(-1)^{k+1}\left(\partial_{\mu,l}\right)^2f_{\mathrm{Q},k}\,.
	\label{eq:df_recur}
\end{align}
From \eqref{eq:df_recur}, the Taylor series of $f_{{\mathrm{Q},k}}(\theta_{\mu l}+\theta_0;\rvec{x})$ can be reduced to a finite linear combination of the zeroth-, first- and second-order derivatives inasmuch as
\begin{align}
	f_{\mathrm{Q},k}(\theta_{\mu l}+\theta_0;\rvec{x})=&\,f_{\mathrm{Q},k}(\theta_{\mu l};\rvec{x})+\sin\theta_0\,\partial_{\mu,l}f_{\mathrm{Q},k}(\theta_{\mu l};\rvec{x})\nonumber\\
	&\,+(1-\cos\theta_0)\,(\partial_{\mu,l})^2f_{\mathrm{Q},k}(\theta_{\mu l};\rvec{x})\,.
	\label{eq:trans}
\end{align}

\subsection{Conditions in gradient and Hessian averaging}
\label{subapp:premise}

\begin{figure}[t]
	\centering
	\includegraphics[width=1\columnwidth]{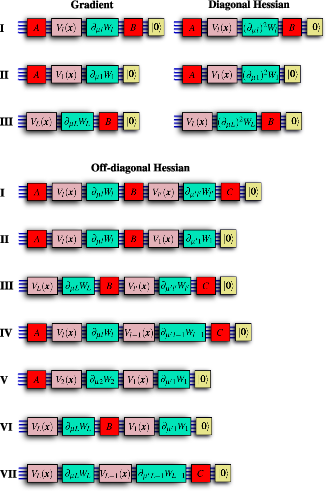}
	\caption{\label{fig:cases}A list of practically-occurring cases in VQAs. All (red) modules $A$, $B$ and $C$ contain two-design trainable and arbitrary nontrainable circuits. In all cases, it is reasonable to require that the \emph{entire} trainable portion of the quantum circuit should be sufficiently deep, such that there exists at least one two-design module that is free from gradient operations at any estimation instance. The TDS condition, thus, corresponds to Case~I for every component type.} 
\end{figure}

Given a trainable module $W_l$, the gradient operation $\partial_{\mu,l}W_l=W_l^{(1)}\sigma_{\mu l} W_l^{(2)}$ can introduce a single-qubit Pauli operator $\sigma_{\mu l}$ associated to the parameter $\theta_{\mu l}$ that divides $W_l$ into subcircuits of unitary operators $W_l^{(1)}$ and $W_l^{(2)}$ that may or may not be two-designs depending on whether these subcircuits are themselves sufficiently deep. Hence, strictly speaking, the details of the circuit averaging procedure would depend on the location of the gradient operations taken. We shall explicitly state the premise in analyzing gradient and Hessian estimation methods:
\begin{enumerate}
	\item In Fig.~\ref{fig:cases}, we list down all the physical cases in which there exists at least an approximate two-design module that is free from gradient operations. The reasonable assumption that the entire circuit should be at least deep enough for the above requirement to hold shall allow us to subsequently analyze sampling errors. 
	\item With such an extent of generality, an exact expression of the MSE for either a gradient or Hessian component is obtained when there exist at least two two-design-approximable training modules sandwiching every training module on which a derivative operation is performed for that component---the \emph{two-design sandwiching}~(TDS) condition~(or {\bf Case~I} in Fig.~\ref{fig:cases}). 
	\item Whenever the TDS condition does not apply for any of the derivative operation in a component, an upper bound of the corresponding MSE is derived.
\end{enumerate}

\subsection{General difference gradient and Hessian method}
\label{subapp:GD}

Equations~\ref{eq:GD1_est} and \ref{eq:GD2_est} are respectively equivalent to
\begin{align}
	\mathrm{GDgrad}^{J,\epsilon}_{\mu,l}f_{\mathrm{Q},k}\equiv&\,\sum^J_{j=1}c_j\,\sinc(j\epsilon/2)\,\partial_{\mu l}f_{\mathrm{Q},k}\,,\label{eq:GD1_est_v2}\\
	\mathrm{GDHess}^{J,\epsilon}_{\mu,l;\mu',l'}f_{\mathrm{Q},k}	\equiv&\,\sum^J_{j=1}c_j[\sinc(j\epsilon/2)]^2\partial_{\mu,l}\partial_{\mu',l'}f_{\mathrm{Q},k}\,.\label{eq:GD2_est_v2}
\end{align}
Similar arguments for the TDS condition in Sec.~\ref{subsec:FD} leads to the following quadratic forms for PEPQCs:
\begin{align}
	\MSE{\mathrm{GD}}{\,\cdot\,}=&\,\rvec{c}^\top(\dyadic{M}_{\displaystyle\,\cdot\,})\,\rvec{c}\,,\,\,\dyadic{M}_{\displaystyle\,\cdot}>\dyadic{0}\,,\nonumber\\
	\dyadic{M}_{\partial f_\mathrm{Q}}=&\,\dfrac{4Jd}{N_\mathrm{T}(d+1)\epsilon^2}\dyadic{A}_{\partial f_\mathrm{Q}}+\dfrac{d^2\,\rvec{v}_1\rvec{v}_1^\top}{2(d+1)(d^2-1)}\,,\nonumber\\
	\dyadic{M}_{\partial\partial f_\mathrm{Q}}=&\,\dfrac{(2J+1)d}{N_\mathrm{T}(d+1)\epsilon^4}\dyadic{A}_{\partial\partial f_\mathrm{Q}}+\dfrac{d^2\,\rvec{v}_2\rvec{v}_2^\top}{2(d+1)(d^2-1)}\,,\nonumber\\
	&\,\qquad\qquad\quad\,\,\,\,\,{\text{(diagonal Hessian components)}}\nonumber\\
	\dyadic{M}_{\partial\partial' f_\mathrm{Q}}=&\,\dfrac{16Jd}{N_\mathrm{T}(d+1)\epsilon^4}\dyadic{A}_{\partial\partial' f_\mathrm{Q}}+\dfrac{d^4\,\rvec{v}_2\rvec{v}_2^\top}{4(d+1)(d^2-1)^2}\,,\nonumber\\
	&\,\qquad\qquad\,\,{\text{(off-diagonal Hessian components)}}
	\label{eq:GD_MSE}
\end{align}
where
\begin{align}
	\rvec{v}_1=\onevec-&\begin{pmatrix}
		\sinc(\epsilon/2)\\
		\sinc(\epsilon)\\
		\vdots\\
		\sinc(J\epsilon/2)
	\end{pmatrix}\,,\,\,\rvec{v}_2=\onevec-\begin{pmatrix}
		[\sinc(\epsilon/2)]^2\\
		[\sinc(\epsilon)]^2\\
		\vdots\\
		[\sinc(J\epsilon/2)]^2
	\end{pmatrix}\,,\nonumber\\
	\dyadic{A}_{\partial f_\mathrm{Q}}=&\,\begin{pmatrix}
		1 & 0 & \cdots & 0\\
		0 & \frac{1}{2^2} & \cdots & 0\\
		\vdots & \vdots & \ddots & \vdots\\
		0 & 0 & \cdots & \frac{1}{J^2}
	\end{pmatrix}\,,\,\,\dyadic{A}_{\partial\partial' f_\mathrm{Q}}=\dyadic{A}_{\partial f_\mathrm{Q}}^2\,,\nonumber\\
	\dyadic{A}_{\partial\partial f_\mathrm{Q}}=&\,2\begin{pmatrix}
		1 & 0 & \cdots & 0\\
		0 & \frac{1}{2^4} & \cdots & 0\\
		\vdots & \vdots & \ddots & \vdots\\
		0 & 0 & \cdots & \frac{1}{J^4}
	\end{pmatrix}+\begin{pmatrix}
		1 \\
		\frac{1}{2^2}\\
		\vdots\\
		\frac{1}{J^2}
	\end{pmatrix}4\begin{pmatrix}
		1\,\,\frac{1}{2^2}\,\,\frac{1}{J^2}
	\end{pmatrix}\,.
	\label{eq:GD_MSE2}
\end{align}
Here, $N_\mathrm{T}$ is still the total number of sampling copies distributed equally to all sampled quantum-circuit functions for one GD approximator per circuit-observable basis operator. Just like the optimal FD~strategy, the optimal GD~strategy in estimating gradient and Hessian components by invoking either \eqref{eq:GD1_est_v2} or \eqref{eq:GD2_est_v2} would entail the choices of \emph{both} $\epsilon$ and normalized $\rvec{c}$ that minimize the relevant operational $\mathrm{MSE}_\mathrm{GD}$ listed in \eqref{eq:GD_MSE}.

\section{Quantum-circuit averages $\left<\,\,\cdot\,\,\right>$}
\label{app:QCavg}

\subsection{Inner-product average of translated $f_{\mathrm{Q},k}$s}
\label{app:avgfQ}

As a warm-up for the upcoming expedition, we calculate the inner product $\MEAN{f_{{\mathrm{Q},k}}(\theta_{\mu l}+\theta_0;\rvec{x}) f_{{\mathrm{Q},k}}(\theta_{\mu l}+\theta'_0;\rvec{x})}{}$ for arbitrary translations $\theta_0$ and $\theta_0'$ on the same randomized parameter~$\theta_{\mu l}$. From Eq.~\eqref{eq:trans},
\begin{align}
	&\,\MEAN{f_{{\mathrm{Q},k}}(\theta_{\mu l}+\theta_0;\rvec{x}) f_{{\mathrm{Q},k}}(\theta_{\mu l}+\theta'_0;\rvec{x})}{}\nonumber\\
	=&\,\MEAN{f_{{\mathrm{Q},k}}^2}{}+(2-\cos\theta_0-\cos\theta'_0)\MEAN{f_{\mathrm{Q},k}(\partial_{\mu,l})^2f_{{\mathrm{Q},k}}}{}\nonumber\\
	&\,+\sin\theta_0\,\sin\theta'_0\MEAN{(\partial_{\mu,l}f_{{\mathrm{Q},k}})^2}{}\nonumber\\
	&\,+(1-\cos\theta_0)(1-\cos\theta'_0)\MEAN{[(\partial_{\mu,l})^2f_{{\mathrm{Q},k}}]^2}{}\,,
\end{align}
where we shall show that $\MEAN{\partial_{\mu,l}f_{\mathrm{Q},k}\,(\partial_{\mu,l})^2f_{\mathrm{Q},k}}{}$ and $\MEAN{f_{\mathrm{Q},k}\,\partial_{\mu,l}f_{\mathrm{Q},k}}{}$ are zero for all three cases shown in Fig.~\ref{fig:cases}.

\begin{center}
	\rule{0.1\textwidth}{1pt}\\[-2.7ex]
	\rule{0.2\textwidth}{1pt}\\[-2.7ex]
	\rule{0.35\textwidth}{1pt}\\[-2.7ex]
	\rule{0.2\textwidth}{1pt}\\[-2.7ex]
	\rule{0.1\textwidth}{1pt}
\end{center}

\begin{flushleft}
	\boxed{\MEAN{f_{\mathrm{Q},k}\,\partial_{\mu,l}f_{\mathrm{Q},k}}{}\text{for {\bf Cases I}}\text{ and {\bf II}}}
\end{flushleft}

Upon taking the average over $W_{L}$ using Eq.~\eqref{eq:pauliV2}, we have 
\begin{align}
	&\,\MEAN{f_{\mathrm{Q},k}\,\partial_{\mu,l}f_{\mathrm{Q},k}}{}\nonumber\\
	=&\,\dfrac{\I}{2(d+1)}\opinner{\rvec{0}}{B^{\dag\otimes2} W^{(2)\dag\otimes2}_l1\otimes\sigma_\mu W^{(2)\otimes2}_lB^{\otimes2}}{\rvec{0}}\nonumber\\
	&\,+\mathrm{c.c.}
\end{align}
Since $W^{(2)\dag\otimes2}_l1\otimes\sigma_\mu W^{(2)\otimes2}_l$ is yet another Pauli operator, these resulting expectation values are real regardless of whether $l=1$ or not (that is, whether $B=1$ correspondingly or not), so that $\MEAN{f_{\mathrm{Q},k}\,\partial_{\mu,l}f_{\mathrm{Q},k}}{\mathrm{I,II}}=0$ for both cases.

\begin{flushleft}
	\boxed{\MEAN{f_{\mathrm{Q},k}\,\partial_{\mu,l}f_{\mathrm{Q},k}}{}\text{for {\bf Case III}}}
\end{flushleft}

For this case, we define the operators
\begin{align}
	\sigma'= &\,W^{(1)\dag}_LV_L^{\dag} O_k\, V_LW^{(1)}_L\,,\label{eq:sigprime}\\
	Q_{\mathrm{III}}=&\,W^{(2)\dag\otimes2}_L\sigma'\otimes\sigma_{\mu L}\sigma'\,W^{(2)\otimes2}_L\,.
	\label{eq:QIII}
\end{align}
From the realization that $\sigma'$ in Eq.~\eqref{eq:sigprime} is a Pauli operator, the following two trace properties
\begin{align}
	\tr{\Uenc{L-1}^{\dag\otimes2} Q_{\mathrm{III}}\,\Uenc{L-1}^{\otimes2}}
	=&\,\tr{\sigma'\otimes\sigma_{\mu L}\sigma'}=0\,,\nonumber\\
	\tr{\Uenc{L-1}^{\dag\otimes2} Q_{\mathrm{III}}\,\Uenc{L-1}^{\otimes2}\tau}
	=&\,\tr{\sigma'\sigma_{\mu L}\sigma'}=\tr{\sigma_{\mu L}}=0
\end{align} 
become apparent, giving us $\MEAN{f_{\mathrm{Q},k}\,\partial_{\mu,l}f_{\mathrm{Q},k}}{\mathrm{III}}=0$.
\begin{center}
	\vspace{-2ex}
	\rule{0.1\textwidth}{1pt}\\[-2.7ex]
	\rule{0.2\textwidth}{1pt}\\[-2.7ex]
	\rule{0.35\textwidth}{1pt}\\[-2.7ex]
	\rule{0.2\textwidth}{1pt}\\[-2.7ex]
	\rule{0.1\textwidth}{1pt}\\[4ex]
\end{center}

By repeating the above calculations, we also find that $\MEAN{(\partial_{\mu,l}f_{\mathrm{Q},k})\,[(\partial_{\mu,l})^2f_{\mathrm{Q},k}]}{\mathrm{I,II,III}}=0$. These results are consistent with the property that average inner products of odd combined derivative order is always zero, another inherent trait from a Pauli-type observable $O_k$.

We are now left with $\MEAN{f_{\mathrm{Q},k}(\partial_{\mu,l})^2f_{{\mathrm{Q},k}}}{}$, $\MEAN{|\partial_{\mu,l}f_{{\mathrm{Q},k}}|^2}{}$ and $\MEAN{|(\partial_{\mu,l})^2f_{{\mathrm{Q},k}}|^2}{}$. First, the average of
\begin{align}
	&\,f_{\mathrm{Q},k}\left(\partial_{\mu,l}\right)^2f_{\mathrm{Q},k}=\dfrac{1}{2}\bra{\rvec{0}}B^{\dag\otimes2} W^{(2)\dag\otimes2}_l1\otimes\sigma_{\mu l} W^{(1)\dag\otimes2}_lV_l^{\dag\otimes2} \nonumber\\
	&\,\times A^{\dag\otimes2} O^{\otimes2}_k A^{\otimes2}V_l^{\otimes2} W^{(1)\otimes2}_l1\otimes\sigma_{\mu l} W^{(2)\otimes2}_lB^{\otimes2}\ket{\rvec{0}}-\dfrac{f^2_{\mathrm{Q},k}}{2}
\end{align}
involves $\MEAN{f^2_{\mathrm{Q},k}}{}=1/(d+1)$ according to Eq.~\eqref{eq:pauliV2}.

\begin{center}
	\rule{0.1\textwidth}{1pt}\\[-2.7ex]
	\rule{0.2\textwidth}{1pt}\\[-2.7ex]
	\rule{0.35\textwidth}{1pt}\\[-2.7ex]
	\rule{0.2\textwidth}{1pt}\\[-2.7ex]
	\rule{0.1\textwidth}{1pt}
\end{center}

\begin{flushleft}
	\boxed{\MEAN{f_{\mathrm{Q},k}\,(\partial_{\mu,l})^2f_{\mathrm{Q},k}}{}\text{for {\bf Case I}}}
\end{flushleft}

With \eqref{eq:pauliV2},
\begin{align}
	&\,\MEAN{f_{\mathrm{Q},k}\,(\partial_{\mu,l})^2f_{\mathrm{Q},k}}{\mathrm{I}}\nonumber\\
	=&\,\dfrac{d}{2(d^2-1)}\left<\opinner{\rvec{0}}{B^{\dag} W^{(2)\dag}_l\sigma_{\mu l} W^{(2)}_lB}{\rvec{0}}^2\right>-\dfrac{d}{2(d^2-1)}\,,
\end{align}
\begin{equation}
	\text{or }\MEAN{f_{\mathrm{Q},k}\,(\partial_{\mu,l})^2f_{\mathrm{Q},k}}{\mathrm{I}}=-\dfrac{d^2}{2(d+1)(d^2-1)}\,.
\end{equation}

\begin{flushleft}
	\boxed{\MEAN{f_{\mathrm{Q},k}\,(\partial_{\mu,l})^2f_{\mathrm{Q},k}}{}\text{for {\bf Case II}}}
\end{flushleft}

We note that 
\begin{align}
	\gamma_\mathrm{II}=\left<\opinner{\rvec{0}}{W^{(2)\dag}_1\sigma_{\mu1} W^{(2)}_1}{\rvec{0}}^2\right>\leq1\,,
	\label{eq:gammaII}
\end{align}
so that
\begin{equation}
	\MEAN{f_{\mathrm{Q},k}\,(\partial_{\mu,l})^2f_{\mathrm{Q},k}}{\mathrm{II}}=-\dfrac{d(1-\gamma_\mathrm{II})}{2(d^2-1)}\,.
\end{equation}

\begin{flushleft}
	\boxed{\MEAN{f_{\mathrm{Q},k}\,(\partial_{\mu,l})^2f_{\mathrm{Q},k}}{}\text{for {\bf Case III}}}
\end{flushleft}

For this case, consider the operator
\begin{align}
	Q_{\mathrm{III}}'=&\,W^{(2)\dag\otimes2}_L1\otimes\sigma_{\mu L} \sigma'^{\otimes2}1\otimes\sigma_{\mu L} W^{(2)\otimes2}_L\,.
	\label{eq:Q3prime}
\end{align}
Its trace properties include $\tr{Q'_\mathrm{III}}=0$,
\begin{align}
	\tr{\Uenc{L-1}^{\dag\otimes2}Q'_{\mathrm{III}}\Uenc{L-1}^{\otimes2}\tau}=&\,\tr{\sigma'\sigma_{\mu L} \sigma'\sigma_{\mu L}}=\gamma_\mathrm{III}(\rvec{x})\,.
\end{align}
With these,
\begin{align}
	\MEAN{f_{\mathrm{Q},k}\,(\partial_{\mu,l})^2f_{\mathrm{Q},k}}{\mathrm{III}}=-\dfrac{d-\gamma_\mathrm{III}(\rvec{x})}{2d(d+1)}\,.
\end{align}
\begin{center}
	\vspace{-1ex}
	\rule{0.1\textwidth}{1pt}\\[-2.7ex]
	\rule{0.2\textwidth}{1pt}\\[-2.7ex]
	\rule{0.35\textwidth}{1pt}\\[-2.7ex]
	\rule{0.2\textwidth}{1pt}\\[-2.7ex]
	\rule{0.1\textwidth}{1pt}\\[4ex]
\end{center}

We catalog the calculations of $\MEAN{(\partial_{\mu,l}f_{{\mathrm{Q},k}})^2}{}$ and $\MEAN{[(\partial_{\mu,l})^2f_{{\mathrm{Q},k}}]^2}{}$ in the following subsections, and simply list the final answers:
\begin{align}	
	&\,\MEAN{f_{{\mathrm{Q},k}}(\theta_{\mu l}+\theta_0;\rvec{x}) f_{{\mathrm{Q},k}}(\theta_{\mu l}+\theta'_0;\rvec{x})}{}\nonumber\\
	=&\,
	\begin{cases}
		\dfrac{d^2\left[\cos\left(\frac{\theta_0-\theta_0'}{2}\right)\right]^2-1}{(d+1)(d^2-1)}	&\mathrm{for}~\textbf{Case I}\,,\\[2ex]
		\dfrac{d(1-\gamma_\mathrm{II})\left[\cos\left(\frac{\theta_0-\theta_0'}{2}\right)\right]^2+d\gamma_\mathrm{II}-1}{d^2-1}	&\mathrm{for}~\textbf{Case II}\,,\\[2ex]
		\dfrac{[d-\overline{\gamma_\mathrm{III}(\rvec{x})}]\left[\cos\left(\frac{\theta_0-\theta_0'}{2}\right)\right]^2+\overline{\gamma_\mathrm{III}(\rvec{x})}}{d(d+1)}	&\mathrm{for}~\textbf{Case III}\,.
	\end{cases}	
\end{align}
By taking $\theta_0=\theta_0'$, we recover the special case $\MEAN{f_{{\mathrm{Q},k}}(\theta_{\mu l}+\theta_0;\rvec{x})^2}{}=1/(d+1)$.

\subsection{Averages of gradient components $\partial f_{\mathrm{Q},k}$}
\label{app:avgdf}

We derive results concerning $\MEAN{\partial_{\mu,l}f_{{\mathrm{Q},k}}}{}$ and $\MEAN{(\partial_{\mu,l}f_{{\mathrm{Q},k}})^2}{}$ for all the three cases in Fig.~\ref{fig:cases}, beginning with the former.

\begin{center}
	\rule{0.1\textwidth}{1pt}\\[-2.7ex]
	\rule{0.2\textwidth}{1pt}\\[-2.7ex]
	\rule{0.35\textwidth}{1pt}\\[-2.7ex]
	\rule{0.2\textwidth}{1pt}\\[-2.7ex]
	\rule{0.1\textwidth}{1pt}
\end{center}

\begin{flushleft}
	\boxed{\MEAN{\partial_{\mu,l}f_\mathrm{Q,k}}{}\text{for {\bf Cases I}}\text{ and {\bf II}}}
\end{flushleft}

Again, as $\Haar{W_L^{\dag}V_L^{\dag}\, O_k\,V_LW_L}=0$ for a Pauli $O_k$,
\begin{align}
	\MEAN{\partial_{\mu,l}f_\mathrm{Q,k}}{\mathrm{I,II}}=0\,.
\end{align}

\begin{flushleft}
	\boxed{\MEAN{\partial_{\mu,l}f_\mathrm{Q,k}}{}\text{for {\bf Case III}}}
\end{flushleft}

We inspect the operator 
\begin{equation}
	Q_{\mathrm{III}}''=W^{(2)\dag}_L\sigma_{\mu L} W^{(1)\dag}_LV_L^{\dag} O_k\, V_LW^{(1)}_LW^{(2)}_L\,.
\end{equation}
Note that $\tr{\sigma_{\mu L} \sigma'}$ is clearly real, so that
\begin{equation}
	\MEAN{\partial_{\mu,L}f_\mathrm{Q,k}}{\mathrm{III}}=\dfrac{\I}{2d}\left<\tr{\sigma_{\mu l}\sigma'}\right>-\dfrac{\I}{2d}\left<\tr{\sigma_{\mu l}\sigma'}\right>=0\,.
\end{equation}

\begin{center}
	\vspace{-3ex}
	\rule{0.1\textwidth}{1pt}\\[-2.7ex]
	\rule{0.2\textwidth}{1pt}\\[-2.7ex]
	\rule{0.35\textwidth}{1pt}\\[-2.7ex]
	\rule{0.2\textwidth}{1pt}\\[-2.7ex]
	\rule{0.1\textwidth}{1pt}\\[4ex]
\end{center}

Now, for the latter:

\begin{center}
	\rule{0.1\textwidth}{1pt}\\[-2.7ex]
	\rule{0.2\textwidth}{1pt}\\[-2.7ex]
	\rule{0.35\textwidth}{1pt}\\[-2.7ex]
	\rule{0.2\textwidth}{1pt}\\[-2.7ex]
	\rule{0.1\textwidth}{1pt}
\end{center}

\begin{flushleft}
	\boxed{\MEAN{|\partial_{\mu,l}f_\mathrm{Q,k}|^2}{}\text{for {\bf Case I}}}
\end{flushleft}

Averaging over $W_L$ yields 
\begin{align}		
	&\,\MEAN{|\partial_{\mu,l}f_\mathrm{Q,k}|^2}{}\nonumber\\
	=&\,-\dfrac{d}{4(d^2-1)}\left<\opinner{\rvec{0}}{B^{\dag\otimes2} W^{(2)\dag\otimes2}_l\sigma_{\mu l}^{\otimes2}W^{(2)\otimes2}_l B^{\otimes2}}{\rvec{0}}\right>\nonumber\\
	&\,+\dfrac{d}{4(d^2-1)}+\mathrm{c.c.}\,,
\end{align}
or
\begin{equation}		
	\MEAN{|\partial_{\mu,l}f_\mathrm{Q,k}|^2}{\mathrm{I}}=\dfrac{d^2}{2(d+1)(d^2-1)}\,.
\end{equation}

\begin{flushleft}
	\boxed{\MEAN{|\partial_{\mu,l}f_\mathrm{Q,k}|^2}{}\text{for {\bf Case II}}}
\end{flushleft}

From \eqref{eq:gammaII}, we simply obtain
\begin{equation}		
	\MEAN{|\partial_{\mu,1}f_\mathrm{Q,k}|^2}{\mathrm{II}}=\dfrac{d(1-\gamma_\mathrm{II})}{2(d^2-1)}\leq\dfrac{d}{2(d^2-1)}\,.
\end{equation}

\begin{flushleft}
	\boxed{\MEAN{|\partial_{\mu,l}f_\mathrm{Q,k}|^2}{}\text{for {\bf Case III}}}
\end{flushleft}

For this case, properties of the operators

\begin{align}
	Q'''_{\mathrm{III(a)}}=&\,W^{(2)\dag\otimes2}_L\sigma_{\mu L}^{\otimes2} \sigma'^{\otimes2}\,W^{(2)\otimes2}_L\,,\nonumber\\
	Q'''_{\mathrm{III(b)}}=&\,W^{(2)\dag\otimes2}_L1\otimes\sigma_{\mu L} \sigma'^{\otimes2}\sigma_{\mu L}\otimes1W^{(2)\otimes2}_L
\end{align}
are necessary, where we recall $\sigma'$ from Eq.~\eqref{eq:sigprime}. To start off,
\begin{align}
	\tr{Q'''_{\mathrm{III(a)}}}=&\,\tr{\sigma_{\mu L}\sigma'}^2=\tr{Q'''_{\mathrm{III(b)}}}\,.
\end{align}
For the trace properties with the swap operator, they are
\begin{align}
	\tr{\Uenc{L-1}^{\dag\otimes2}Q'''_{\mathrm{III(a)}}\Uenc{L-1}^{\otimes2}\tau}=&\,\tr{\sigma_{\mu L}\sigma'\sigma_{\mu L}\sigma'}=\gamma_\mathrm{III}(\rvec{x})\,,\nonumber\\
	\tr{\Uenc{L-1}^{\dag\otimes2}Q'''_{\mathrm{III(b)}}\Uenc{L-1}^{\otimes2}\tau}=&\,\tr{\sigma'\sigma_{\mu L} \sigma_{\mu L}\sigma'}=d\,.
\end{align}
These are critical in evaluating the average over $W_{L-1}$ by invoking Eq.~\eqref{eq:pauliV2}:
\begin{align}
	\gamma_\mathrm{III}(\rvec{x})\equiv&\,\left<\tr{(\sigma_{\mu L}\sigma')^2}\right>\,,\nonumber\\
	\gamma'_\mathrm{III}(\rvec{x})\equiv&\,\left<\tr{\sigma_{\mu L}\sigma'}^2\right>\,,\nonumber\\	
	\MEAN{|\partial_{\mu,L}f_\mathrm{Q,k}|^2}{\mathrm{III}}
	=&\,-\dfrac{\gamma'_\mathrm{III}(\rvec{x})+\gamma_\mathrm{III}(\rvec{x})}{2d(d+1)}+\dfrac{\gamma'_\mathrm{III}(\rvec{x})+d}{2d(d+1)}\nonumber\\
	=&\,\dfrac{d-\gamma_\mathrm{III}(\rvec{x})}{2d(d+1)}\leq\dfrac{1}{d+1}\,.
	\label{eq:gammaIII}
\end{align}
The final inequality is of the Cauchy--Schwarz type
\begin{align}
	\tr{[\sigma_{\mu l}\sigma']^2}^2\leq\tr{\sigma_{\mu l}^2}\tr{[\sigma'\sigma_{\mu l}\sigma']^2}=d^2\,,
\end{align}
or $-d\leq\gamma_\mathrm{III}(\rvec{x})\leq d$.

\begin{center}
	\vspace{-2ex}
	\rule{0.1\textwidth}{1pt}\\[-2.7ex]
	\rule{0.2\textwidth}{1pt}\\[-2.7ex]
	\rule{0.35\textwidth}{1pt}\\[-2.7ex]
	\rule{0.2\textwidth}{1pt}\\[-2.7ex]
	\rule{0.1\textwidth}{1pt}\\[4ex]
\end{center}

\subsection{Averages of diagonal Hessian components $\partial\partial f_{\mathrm{Q},k}$}
\label{app:avgd2f}

From \eqref{eq:df,d2f}, just like $\partial f_{\mathrm{Q},k}=0$, it can be easily confirmed that $\MEAN{\partial\partial f_{\mathrm{Q},k}}{\mathrm{I,II,III}}=0$. For the squared averages, since

\begin{align}
	&\,\left[\left(\partial_{\mu,l}\right)^2f_\mathrm{Q,k}\right]^2=\dfrac{1}{4}\left[f_\mathrm{Q,k}^2+D_\mathrm{I}-2\,D_\mathrm{II}\right]\,,\nonumber\\	
	&\,\,D_\mathrm{I}=\opinner{\rvec{0}}{B^{\dag\otimes2} W^{(2)\dag\otimes2}_l\sigma_{\mu l}^{\otimes2} W^{(1)\dag\otimes2}_lV_l^{\dag\otimes2} \nonumber\\
	&\,\qquad\quad\,\,\,\times A^{\dag\otimes2} O^{\otimes2} A^{\otimes2}V_l^{\otimes2}W^{(1)\otimes2}_l\sigma_{\mu l}^{\otimes2} W^{(2)\otimes2}_lB^{\otimes2}}{\rvec{0}}\,,\nonumber\\
	&\,D_{\mathrm{II}}=\bra{\rvec{0}}B^{\dag\otimes2} W^{(2)\dag\otimes2}_l1\otimes\sigma_{\mu l} W^{(1)\dag\otimes2}_lV_l^{\dag\otimes2} \nonumber\\
	&\,\qquad\,\,\,\times A^{\dag\otimes2} O^{\otimes2}_k A^{\otimes2}V_l^{\otimes2} W^{(1)\otimes2}_l1\otimes\sigma_{\mu l} W^{(2)\otimes2}_lB^{\otimes2}\ket{\rvec{0}}\,,
\end{align}
we need the averages of $D_{\mathrm{I}}$ and $D_{\mathrm{II}}$ in all three cases.

\begin{center}
	\rule{0.1\textwidth}{1pt}\\[-2.7ex]
	\rule{0.2\textwidth}{1pt}\\[-2.7ex]
	\rule{0.35\textwidth}{1pt}\\[-2.7ex]
	\rule{0.2\textwidth}{1pt}\\[-2.7ex]
	\rule{0.1\textwidth}{1pt}
\end{center}

\begin{flushleft}
	\boxed{\MEAN{|(\partial_{\mu,l})^2f_\mathrm{Q,k}|^2}{}\text{for {\bf Case I}}}
\end{flushleft}

That $\MEAN{D_\mathrm{I}}{}=1/(d+1)$ is immediate. On the other hand, 
\begin{align}
	\MEAN{D_\mathrm{II}}{}=&\,\dfrac{d}{d^2-1}\left<\opinner{\rvec{0}}{B^{\dag\otimes2} W^{(2)\dag\otimes2}_l\sigma_{\mu l}^{\otimes2} W^{(2)\otimes2}_lB^{\otimes2}}{\rvec{0}}\right>\nonumber\\
	&\,-\dfrac{1}{d^2-1}\nonumber\\
	=&\,-\dfrac{1}{(d+1)(d^2-1)}\,.
\end{align}
Altogether,
\begin{equation}
	\MEAN{|(\partial_{\mu,l})^2f_\mathrm{Q,k}|^2}{\mathrm{I}}=\dfrac{d^2}{2(d+1)(d^2-1)}\,.
\end{equation}

\begin{flushleft}
	\boxed{\MEAN{|(\partial_{\mu,l})^2f_\mathrm{Q,k}|^2}{}\text{for {\bf Case II}}}
\end{flushleft}

If $l=1$, then $\MEAN{D_\mathrm{I}}{}=1/(d+1)$ still, but
\begin{equation}
	\MEAN{D_\mathrm{II}}{}=\dfrac{d\gamma_\mathrm{II}-1}{d^2-1}\,,
\end{equation}
so that
\begin{equation}
	\MEAN{|(\partial_{\mu,1})^2f_\mathrm{Q,k}|^2}{\mathrm{II}}=\dfrac{d-d\gamma_\mathrm{II}}{2(d^2-1)}\leq\dfrac{d}{2(d^2-1)}\,,
\end{equation}
where $\gamma_\mathrm{II}$ is as defined in \eqref{eq:gammaII}.

\begin{flushleft}
	\boxed{\MEAN{|(\partial_{\mu,l})^2f_\mathrm{Q,k}|^2}{}\text{for {\bf Case III}}}
\end{flushleft}

From \eqref{eq:Q3prime} and the operator definition
\begin{align}
	Q''''_\mathrm{III}=&\,W^{(2)\dag\otimes2}_L\sigma_{\mu L}^{\otimes2} \sigma'^{\otimes2}\sigma_{\mu L}^{\otimes2} W^{(2)\otimes2}_L\,,
\end{align}
whose trace properties include $\tr{Q''''_\mathrm{III}}=0$ and
\begin{equation}
	\tr{\Uenc{L-1}^{\dag\otimes2}Q''''_{\mathrm{III}}\Uenc{L-1}^{\otimes2}\tau}=\tr{\sigma_{\mu L}\sigma'\sigma'\sigma_{\mu L}}=d\,.
\end{equation}
With these,
\begin{align}
	\MEAN{D_\mathrm{I}}{}=&\,\MEAN{\opinner{\rvec{0}}{B^{\dag\otimes2}\,Q''''_\mathrm{III}B^{\otimes2}}{\rvec{0}}}{}=\dfrac{1}{d+1}\,,\nonumber\\
	\MEAN{D_\mathrm{II}}{}=&\,\MEAN{\opinner{\rvec{0}}{B^{\dag\otimes2}\,Q'_\mathrm{III}B^{\otimes2}}{\rvec{0}}}{}=\dfrac{\gamma_\mathrm{III}(\rvec{x})}{d(d+1)}\,,
\end{align}
which finally brings us to
\begin{equation}
	\MEAN{|(\partial_{\mu,L})^2f_\mathrm{Q,k}|^2}{\mathrm{III}}=\dfrac{d-\gamma_\mathrm{III}(\rvec{x})}{2d(d+1)}\leq\dfrac{1}{d+1}\,.
\end{equation}

\begin{center}
	\rule{0.1\textwidth}{1pt}\\[-2.7ex]
	\rule{0.2\textwidth}{1pt}\\[-2.7ex]
	\rule{0.35\textwidth}{1pt}\\[-2.7ex]
	\rule{0.2\textwidth}{1pt}\\[-2.7ex]
	\rule{0.1\textwidth}{1pt}\\[4ex]
\end{center}

\subsection{Averages of off-diagonal Hessian components $\partial_{\mu,l}\partial_{\mu',l'} f_{\mathrm{Q},k}$}
\label{app:avgddf}

The averages $|\partial_{\mu,l}\partial_{\mu',l'} f_{\mathrm{Q},k}|^2$ are to be computed for a total of seven cases depicted in Fig.~\ref{fig:cases}. Each component consists of a summation of four pieces:
\begin{align}
	&\,\partial_{\mu,l}\partial_{\mu',l'} f_{\mathrm{Q},k}=[(1)+(1)_{\mathrm{c.c.}}+(2)+(2)_{\mathrm{c.c.}}]\,,\nonumber\\
	&\,(1)=-\frac{1}{4}\bra{\rvec{0}}C^\dag W^{(2)\dag}_{l'}\sigma_{\mu'l'}W^{(1)\dag}_{l'}V^\dag_{l'}B^\dag W^{(2)\dag}_{l}\sigma_{\mu l}W^{(1)\dag}_{l}V^\dag_l\nonumber\\
	&\,\qquad\qquad\quad\times A^\dag O_k AV_lW_lBV_{l'}W_{l'}C\ket{\rvec{0}}\,,\nonumber\\
	&\,(2)=\frac{1}{4}\bra{\rvec{0}}C^\dag W^{\dag}_{l'}V^\dag_{l'}B^\dag W^{(2)\dag}_{l}\sigma_{\mu l}W^{(1)\dag}_{l}V^\dag_lA^\dag O_k A\nonumber\\
	&\,\qquad\qquad\,\times V_lW_lBV_{l'}^\dag W^{(1)\dag}_{l'}\sigma_{\mu' l'}W^{(2)\dag}_{l'}C\ket{\rvec{0}}\,.
\end{align}
Its squared average thus involves 16 terms. We shall list all of their combinations conveniently for every case.

\begin{center}
	\rule{0.1\textwidth}{1pt}\\[-2.7ex]
	\rule{0.2\textwidth}{1pt}\\[-2.7ex]
	\rule{0.35\textwidth}{1pt}\\[-2.7ex]
	\rule{0.2\textwidth}{1pt}\\[-2.7ex]
	\rule{0.1\textwidth}{1pt}
\end{center}

\begin{flushleft}
	\boxed{\MEAN{|\partial_{\mu,l}\partial_{\mu',l'}f_\mathrm{Q,k}|^2}{}\text{for {\bf Cases I} and {\bf IV}}}
\end{flushleft}

Let $\widetilde\sigma_{\mu l}=W^{(1)\dag}_{l'}V_{l'}^\dag B^\dag W^{(2)\dag}_{l}\sigma_{\mu l}W^{(2)}_lBV_{l'}W^{(1)}$. Then upon denoting $\widetilde{\gamma}=\tr{(\sigma_{\mu'l'}\widetilde{\sigma}_{\mu l})^2}$ and $\widetilde{\gamma}'=\tr{\sigma_{\mu'l'}\widetilde{\sigma}_{\mu l}}^2$, we get

\begin{align}
	\MEAN{(1)^2+(1)^2_{\mathrm{c.c.}}}{}=&\,\MEAN{(2)^2+(2)^2_{\mathrm{c.c.}}}{}=\dfrac{1}{8d(d+1)^2}\MEAN{\widetilde{\gamma}+\widetilde{\gamma}'}{}\,,\nonumber\\
	2\MEAN{(1)\times(1)_{\mathrm{c.c.}}}{}=&\,\dfrac{d^2+d-1}{8(d+1)(d^2-1)}-\dfrac{\MEAN{\widetilde{\gamma}'}{}}{8d(d+1)(d^2-1)}\,,\nonumber\\
	2\MEAN{(1)\times(2)}{}=&\,\dfrac{\MEAN{\widetilde{\gamma}'-d\widetilde{\gamma}}{}}{8d(d+1)(d^2-1)}+\dfrac{1}{8(d+1)(d^2-1)}\,,\nonumber\\
	2\MEAN{(1)\times(2)_{\mathrm{c.c.}}}{}=&\,\dfrac{\MEAN{\widetilde{\gamma}+\widetilde{\gamma}'}{}}{8d(d+1)(d^2-1)}-\dfrac{d}{8(d+1)(d^2-1)}\,,\nonumber\\
	2\MEAN{(2)\times(2)_{\mathrm{c.c.}}}{}=&\,2\MEAN{(1)\times(1)_{\mathrm{c.c.}}}{}\,.
\end{align}
These amount to
\begin{align}
	\MEAN{|\partial_{\mu,l}\partial_{\mu',l'}f_\mathrm{Q,k}|^2}{\mathrm{IV}}=&\,\dfrac{d^2+\MEAN{\widetilde{\gamma}'}{}}{4(d+1)(d^2-1)}\nonumber\\
	&\,\leq \dfrac{d^2}{2(d+1)(d^2-1)}\,.
\end{align}

In {\bf Case~I}, $\MEAN{\widetilde{\gamma}'}{}=d^2/(d^2-1)$, so that
\begin{equation}
	\MEAN{|\partial_{\mu,l}\partial_{\mu',l'}f_\mathrm{Q,k}|^2}{\mathrm{I}}=\dfrac{d^4}{4(d+1)(d^2-1)^2}\,.
\end{equation}

\begin{flushleft}
	\boxed{\MEAN{|\partial_{\mu,l}\partial_{\mu',l'}f_\mathrm{Q,k}|^2}{}\text{for {\bf Cases II} and {\bf V}}}
\end{flushleft}

Let us define the shorthand notations $\alpha=\opinner{\rvec{0}}{W^{(2)\dag}_{l'}\sigma_{\mu'l'}\widetilde{\sigma}_{\mu l}W^{(2)}_{l'}}{\rvec{0}}$, $\beta=\opinner{\rvec{0}}{W^{(2)\dag\otimes2}_{l'}\sigma_{\mu'l'}\widetilde{\sigma}_{\mu l}\sigma_{\mu'l'}\otimes\widetilde{\sigma}_{\mu l} W^{(2)\otimes2}_{l'}}{\rvec{0}}$ and $\gamma=\opinner{\rvec{0}}{W^{(2)\dag}_{l'}\sigma_{\mu'l'}W^{(2)}_{l'}}{\rvec{0}}$. Then,
\begin{align}
	\MEAN{(1)^2+(1)^2_{\mathrm{c.c.}}}{}=&\,\MEAN{(2)^2+(2)^2_{\mathrm{c.c.}}}{}=\dfrac{\RE{\MEAN{\alpha^2}{}}}{8(d+1)}\,,\nonumber\\
	2\MEAN{(1)\times(1)_{\mathrm{c.c.}}}{}=&\,\dfrac{d}{8(d^2-1)}-\dfrac{\MEAN{|\alpha|^2}{}}{8(d^2-1)}\,,\nonumber\\
	2\MEAN{(1)\times(2)}{}=&\,\dfrac{\MEAN{|\alpha|^2}{}}{8(d^2-1)}-\dfrac{d\MEAN{\beta}{}}{8(d^2-1)}\,,\nonumber\\
	2\MEAN{(1)\times(2)_{\mathrm{c.c.}}}{}=&\,\dfrac{\MEAN{|\alpha|^2}{}}{8(d^2-1)}-\dfrac{d\MEAN{\gamma^2}{}}{8(d^2-1)}\,,\nonumber\\
	2\MEAN{(2)\times(2)_{\mathrm{c.c.}}}{}=&\,2\MEAN{(1)\times(1)_{\mathrm{c.c.}}}{}\,,
\end{align}
such that
\begin{align}
	\MEAN{|\partial_{\mu,l}\partial_{\mu',l'}f_\mathrm{Q,k}|^2}{\mathrm{V}}=&\,\dfrac{d\MEAN{1+\RE{\alpha^2}-\beta-\gamma^2}{}}{4(d^2-1)}\nonumber\\
	&\,\leq \dfrac{3d}{4(d^2-1)}\,.
\end{align}
The inequality is deducible from the basic inequalities $0\leq\gamma^2\leq1$, $-1\leq\beta\leq1$, and $\RE{\alpha^2}\leq |\alpha|^2\leq1$.

When {\bf Case~II} holds, one arrives at $\MEAN{\alpha^2}{}=\MEAN{\gamma^2}{}/(d+1)$ and $\MEAN{\beta}{}=(d\MEAN{\gamma^2}{}-1)/(d^2-1)$, yielding
\begin{equation}
	\MEAN{|\partial_{\mu,l}\partial_{\mu',l'}f_\mathrm{Q,k}|^2}{\mathrm{II}}=\dfrac{d^3(1-\MEAN{\gamma^2}{})}{4(d^2-1)^2}\leq \dfrac{d^3}{4(d^2-1)^2}\,.
\end{equation}

\begin{table*}
	\begin{tabular}[t]{cccc}
		{\bf Case}\quad &\quad $\left<\left|\partial_{\mu,l} f_{\mathrm{Q},k}\right|^2\right>$\quad &\quad $\left<\left|(\partial_{\mu,l})^2 f_{\mathrm{Q},k}\right|^2\right>$\quad &\quad $\left<\left|\partial_{\mu,l} \partial_{\mu',l'} f_{\mathrm{Q},k}\right|^2\right>$\\[1ex]
		\hline\hline\\[-2ex]
		{\bf I}\quad &\quad $\dfrac{d^2}{2(d+1)(d^2-1)}$\quad &\quad $\dfrac{d^2}{2(d+1)(d^2-1)}$\quad &\quad $\dfrac{d^4}{4(d+1)(d^2-1)^2}$\\[2ex]
		{\bf II}\quad &\quad $\leq\dfrac{d}{2(d^2-1)}$\quad &\quad $\leq\dfrac{d}{2(d^2-1)}$\quad &\quad $\leq\dfrac{d^3}{4(d^2-1)^2}$\\[2ex]
		{\bf III}\quad &\quad $\leq\dfrac{1}{d+1}$\quad &\quad $\leq\dfrac{1}{d+1}$\quad &\quad $\leq\dfrac{1}{2(d+1)}$
	\end{tabular}~~~~\begin{tabular}[t]{cc}
		{\bf Case}\quad &\quad $\left<\left|\partial_{\mu,l} \partial_{\mu',l'} f_{{\mathrm{Q},k}}\right|^2\right>$\\[1ex]
		\hline\hline\\[-2ex]
		{\bf IV}\quad &\quad $\leq\dfrac{d^2}{2(d+1)(d^2-1)}$\\[2ex]
		{\bf V}\quad &\quad $\leq\dfrac{3d}{4(d^2-1)}$\\[2ex]
		{\bf VI}\quad &\quad $\leq\dfrac{d}{2(d^2-1)}$\\[2ex]
		{\bf VII}\quad &\quad $\leq\dfrac{1}{2(d+1)}$
	\end{tabular}
	\caption{\label{tab:avg_res}Summary table listing all averages (and their upper bounds) of squared-magnitudes for all types of gradient and Hessian components. We remind the reader that the notation $\partial_{\mu,l} \partial_{\mu',l'}$ refers to the off-diagonal components, where either $\mu\neq\mu'$, or $l\neq l'$, or both.}
\end{table*}

\begin{flushleft}
	\boxed{\MEAN{|\partial_{\mu,l}\partial_{\mu',l'}f_\mathrm{Q,k}|^2}{}\text{for {\bf Cases III} and {\bf VII}}}
\end{flushleft}

For these two cases, let us focus on the Pauli operators
\begin{align}
	\sigma_0\equiv&\, W^{(1)\dag}_lV^\dag_lA^\dag O_kA\,V_lW^{(1)}_l\,,\nonumber\\
	\sigma_1\equiv&\,V^\dag_{l'}B^\dag W^{(2)\dag}_l\sigma_0 W^{(2)}_l BV_{l'}\,,\nonumber\\
	\sigma_2\equiv&\,V^\dag_{l'}B^\dag W^{(2)\dag}_l\sigma_{\mu l} W^{(2)}_l BV_{l'}\,.
\end{align}
In terms of these Pauli operators and the rotated versions $\widetilde{\sigma}_j=W^{(1)\dag}_{l'}\sigma_j W^{(1)}_{l'}$, we define a new set of parameters: $\alpha'=\tr{\sigma_{\mu'l'}\widetilde{\sigma}_2\widetilde{\sigma}_1}$, $\beta'=\tr{(\sigma_{\mu'l'}\widetilde{\sigma}_2\widetilde{\sigma}_1)^2}$. These lead to

\begin{align}
	\MEAN{(1)^2+(1)^2_{\mathrm{c.c.}}}{}=&\,\MEAN{(2)^2+(2)^2_{\mathrm{c.c.}}}{}=\dfrac{\RE{\MEAN{\alpha'^2+\beta'}{}}}{8d(d+1)}\,,\nonumber\\
	2\MEAN{(1)\times(1)_{\mathrm{c.c.}}}{}=&\,\dfrac{\MEAN{|\alpha'|^2}{}+d}{8d(d+1)}\,,\nonumber\\
	2\MEAN{(1)\times(2)}{}=&\,-\dfrac{\MEAN{\alpha'^2+\tr{(\sigma_1\sigma_2)^2}}{}}{8d(d+1)}\,,\nonumber\\
	2\MEAN{(1)\times(2)_{\mathrm{c.c.}}}{}=&\,-\dfrac{\MEAN{|\alpha'|^2+\beta'}{}}{8d(d+1)}\,,\nonumber\\
	2\MEAN{(2)\times(2)_{\mathrm{c.c.}}}{}=&\,2\MEAN{(1)\times(1)_{\mathrm{c.c.}}}{}\,,
\end{align}
giving us
\begin{align}
	\MEAN{|\partial_{\mu,l}\partial_{\mu',l'}f_\mathrm{Q,k}|^2}{\mathrm{VII}}=&\,\dfrac{d-\langle\tr{(\sigma_1\sigma_2)^2}\rangle}{4d(d+1)}\leq\dfrac{1}{2(d+1)}\,,
\end{align}
where we employed the Cauchy--Schwarz inequality $-d\leq\tr{(\sigma_1\sigma_2)^2}\leq d$.

As $\tr{(\sigma_1\sigma_2)^2}$ is independent of $B$, we see that {\bf Case~III} offers no further tightening to the above inequality: 
\begin{equation}
\MEAN{|\partial_{\mu,l}\partial_{\mu',l'}f_\mathrm{Q,k}|^2}{\mathrm{III}}\leq\dfrac{1}{2(d+1)}\,.
\end{equation}

\begin{flushleft}
	\boxed{\MEAN{|\partial_{\mu,l}\partial_{\mu',l'}f_\mathrm{Q,k}|^2}{}\text{for {\bf Cases VI}}}
\end{flushleft}

In this final case, do expressions are
\begin{align}
	\MEAN{(1)^2+(1)^2_{\mathrm{c.c.}}}{}=&\,\MEAN{(2)^2+(2)^2_{\mathrm{c.c.}}}{}=\dfrac{1}{8}\RE{\MEAN{x^2}{}}\,,\nonumber\\
	2\MEAN{(1)\times(1)_{\mathrm{c.c.}}}{}=&\,\dfrac{1}{8}\MEAN{|x|^2}{}\,,\nonumber\\
	2\MEAN{(1)\times(2)}{}=&\,-\dfrac{1}{8}\MEAN{xy}{}\,,\nonumber\\
	2\MEAN{(1)\times(2)_{\mathrm{c.c.}}}{}=&\,-\dfrac{1}{8}\MEAN{xy^*}{}\,,\nonumber\\
	2\MEAN{(2)\times(2)_{\mathrm{c.c.}}}{}=&\,\dfrac{1}{8}\MEAN{|y|^2}{}\,,
\end{align}
with
\begin{align}
	w_{\mu'l'}=&\,\opinner{\rvec{0}}{W^{(2)}_{l'}\sigma_{\mu'l'}W^{(2)}_{l'}}{\rvec{0}}^2\leq1\,,\nonumber\\
	\MEAN{x^2}{}=&\,w_{\mu'l'}\dfrac{\MEAN{\tr{\sigma_0\sigma_{\mu l}}^2+\tr{\sigma_0\sigma_{\mu l}\sigma_0\sigma_{\mu l}}}{}}{d(d+1)}\,,\nonumber\\
	\MEAN{|x|^2}{}=&\,\MEAN{|y|^2}{}\nonumber\\
	=&\,\MEAN{\tr{\sigma_0\sigma_{\mu l}}^2}{}\dfrac{w_{\mu'l'}-1/d}{d(d^2-1)}+\dfrac{d-w_{\mu'l'}}{d^2-1}\,,\nonumber\\
	\MEAN{\RE{xy}}{}=&\,\dfrac{\tr{\sigma_0\sigma_{\mu l}}^2(dw_{\mu'l'}-1)}{d(d^2-1)}\nonumber\\
	&\,+\dfrac{\tr{(\sigma_0\sigma_{\mu l})^2}(d-w_{\mu'l'})}{d(d^2-1)}\,,\nonumber\\
	\MEAN{\RE{xy^*}}{}=&\,\dfrac{w_{\mu'l'}}{d+1}\left(1+\dfrac{\tr{\sigma_0\sigma_{\mu l}}^2}{d}\right)\,.
\end{align}
Combining these expressions nabs us
\begin{align}
	\MEAN{|\partial_{\mu,l}\partial_{\mu',l'}f_\mathrm{Q,k}|^2}{\mathrm{VI}}=&\,\dfrac{(1-w_{\mu'l'})(d-\tr{\sigma_0\sigma_{\mu l}\sigma_0\sigma_{\mu l}})}{4(d^2-1)}\nonumber\\
	\leq&\,\dfrac{d}{2(d^2-1)}\,.
\end{align}

\begin{center}
	\vspace{-1ex}
	\rule{0.1\textwidth}{1pt}\\[-2.7ex]
	\rule{0.2\textwidth}{1pt}\\[-2.7ex]
	\rule{0.35\textwidth}{1pt}\\[-2.7ex]
	\rule{0.2\textwidth}{1pt}\\[-2.7ex]
	\rule{0.1\textwidth}{1pt}\\[4ex]
\end{center}

\subsection{Summary table of all gradient and Hessian averages}
\label{app:summaryallcase}

Without referring to the details in Appendices~\ref{app:avgdf} through \ref{app:avgddf}, Table~\ref{tab:avg_res} summarizes the important averages of squared magnitudes for all gradient- and Hessian-component types. Evidently, all averages are at most $O(1/d)$, which is crucial for revealing the beneficial statistical properties of optimized FD and GD estimators.

\section{Optimized sampling errors for all cases}
\label{app:smp_err}

\subsection{Multinomial sampling distribution}

If each $d$-dimensional Pauli basis operator $O_k$ of the variational quantum circuit is measured independently, then each measurement of a fixed number of sampling copies $N$ is the eigenbasis $\{\ket{o_{kl}}\bra{o_{kl}}\}^{d-1}_{l=0}$ of $O_k$, where relative frequencies $\nu_{kl}(\rvec{\theta};\rvec{x})=n_{kl}(\rvec{\theta};\rvec{x})/N\rightarrow p_{kl}(\rvec{\theta};\rvec{x})$ are recorded. Explicitly, these relative frequencies satisfy the following basic statistical identities:

\begin{align}
	&\,\AVG{}{\nu_{kl}(\rvec{\theta};\rvec{x})}=p_{kl}(\rvec{\theta};\rvec{x})\,,\nonumber\\
	&\,\AVG{}{\nu_{kl}(\rvec{\theta};\rvec{x})\nu_{kl'}(\rvec{\theta}';\rvec{x}')}=(1\!-\!\delta_{\rvec{\theta},\rvec{\theta}'}\delta_{\rvec{x},\rvec{x}'})p_{kl}(\rvec{\theta};\rvec{x})p_{kl'}(\rvec{\theta}';\rvec{x}')\nonumber\\
	&\quad+\dfrac{\delta_{\rvec{\theta},\rvec{\theta}'}\delta_{\rvec{x},\rvec{x}'}}{N}\left[\delta_{l,l'}p_{kl}(\rvec{\theta};\rvec{x})+(N-1)p_{kl}(\rvec{\theta};\rvec{x})p_{kl'}(\rvec{\theta};\rvec{x})\right],\nonumber\\
	&\,\AVG{}{\nu_{kl}(\rvec{\theta};\rvec{x})\nu_{{k'\neq k}\,l'}(\rvec{\theta}';\rvec{x}')}=p_{kl}(\rvec{\theta};\rvec{x})p_{k'l'}(\rvec{\theta}';\rvec{x})\,.
	\label{eq:multinom_prop}
\end{align}
The above identities hold whenever circuits of different training parameters are sampled independently.

\subsection{Function estimation}

We start with the estimation of $f_\mathrm{Q}$ using the unbiased estimator in \eqref{eq:fest}. Together with the multinomial identities in~\eqref{eq:multinom_prop}, this leads to 
\begin{equation}
	\MSE{}{f_\mathrm{Q}}=\dfrac{1}{N_\mathrm{T}}\left(1-\overline{\MEAN{f_{\mathrm{Q},k}^2}{}}\right)\,,
\end{equation}
where $N_\mathrm{T}=N$ is the total number of copies \emph{per basis observable} needed to complete one function estimation evaluated at the circuit parameters $\rvec{\theta}$. For two-design approximable PEPQCs, by making use of Eq.~\eqref{eq:pauliV2}, we quickly find that
\begin{equation}
	\MEAN{f_{\mathrm{Q},k}(\theta_{\mu l};\rvec{x})^2}{}=\MEAN{f_{\mathrm{Q},k}(\theta_{\mu l}+\theta_0;\rvec{x})^2}{}=\dfrac{1}{d+1}
	\label{eq:avgf2}
\end{equation}
for any $\theta_0$~(see Appendix~\ref{app:avgfQ}), so that
\begin{equation}
	\MSE{}{f_\mathrm{Q}}=\dfrac{d}{N_\mathrm{T}(d+1)}\,.
	\label{eq:MSE_f}
\end{equation}

\subsection{Optimally-tuned FD estimators for general cases}
\label{app:FDallcase}

In their most general forms, the FD~MSEs, which are linear combinations of finite-copy and nonzero-$\epsilon$ approximation squared-errors, read

\begin{align}
	\MSE{\mathrm{FD}}{\partial f_\mathrm{Q}}=&\,\dfrac{4d}{N_\mathrm{T}(d+1)\epsilon^2}+\MEAN{(\partial f_{\mathrm{Q}})^2}{}g_1(\epsilon/2)\,,\nonumber\\
	\MSE{\mathrm{FD}}{\partial\partial f_\mathrm{Q}}=&\,\dfrac{18d}{N_\mathrm{T}(d+1)\epsilon^4}+\MEAN{(\partial\partial f_{\mathrm{Q}})^2}{}g_2(\epsilon/2)\,,\nonumber\\
	\MSE{\mathrm{FD}}{\partial\partial' f_\mathrm{Q}}=&\,\dfrac{16d}{N_\mathrm{T}(d+1)\epsilon^4}+\MEAN{(\partial\partial' f_{\mathrm{Q}})^2}{}g_2(\epsilon/2)\,,
	\label{eq:FD_MSE_gen}
\end{align}
with $g_1(\epsilon/2)=[1-\sinc(\epsilon/2)]^2$, $g_2(\epsilon/2)=\{1-[\sinc(\epsilon/2)]^2\}^2$, and all the three average terms $\MEAN{(\partial f_{\mathrm{Q}})^2}{}$,  $\MEAN{(\partial\partial f_{\mathrm{Q}})^2}{}$ and  $\MEAN{(\partial\partial' f_{\mathrm{Q}})^2}{}$ have been rigorously worked out in Appendices~\ref{app:avgdf} through \ref{app:avgddf}.

Keeping in their arbitrary forms, we can proceed to minimize all these MSEs and derive optimal FD estimators in the regime $\epsilon\ll 1$. Using the Taylor approximations $g_1(\epsilon/2)\cong\epsilon^4/576$ and $g_2(\epsilon/2)\cong\epsilon^4/144$, we identify two different functional structures
\begin{align}
	\MSE{\mathrm{FD}}{\partial f_\mathrm{Q}}\,\,\widehat{=}&\,\dfrac{A_1}{\epsilon^2}+\dfrac{\epsilon^4}{576} B_1\,,\nonumber\\
	\MSE{\mathrm{FD}}{\partial\partial f_\mathrm{Q}},\MSE{\mathrm{FD}}{\partial\partial' f_\mathrm{Q}}\,\,\widehat{=}&\,\dfrac{A_2}{\epsilon^4}+\dfrac{\epsilon^4}{144} B_2\,,
\end{align}
where $A_1,A_2,B_1,B_2>0$.

Now, optimizing over $\epsilon$ for both kinds of structures,
\begin{align}
	\min_\epsilon\left\{\dfrac{A_1}{\epsilon^2}+\dfrac{\epsilon^4B_1}{576} \right\}=&\,\left(\dfrac{3A_1^2B_1}{256}\right)^{1/3}\!\!\!\!\text{with }\epsilon=\left(\dfrac{288A_1}{B_1}\right)^{1/6}\!\!,\nonumber\\
	\min_\epsilon\left\{\dfrac{A_2}{\epsilon^4}+\dfrac{\epsilon^4B_2}{144} \right\}=&\,\dfrac{\sqrt{A_2B_2}}{6}\,\,\text{with }\epsilon=\left(\dfrac{144A_2}{B_2}\right)^{1/8}.
	\label{eq:min_twostructs}
\end{align}
These give the complete set of approximate $\epsilon_\mathrm{opt}$s for all types of sampling inasmuch as
\begin{align}
	\epsilon_\mathrm{opt}(\partial f_\mathrm{Q})\cong&\,\left[\dfrac{1152d}{\MEAN{(\partial f_{\mathrm{Q}})^2}{}N_\mathrm{T}(d+1)}\right]^{1/6}\,,\nonumber\\
	\epsilon_\mathrm{opt}(\partial\partial f_\mathrm{Q})\cong&\,\left[\dfrac{2592d}{\MEAN{(\partial\partial f_{\mathrm{Q}})^2}{}N_\mathrm{T}(d+1)}\right]^{1/8}\,,\nonumber\\
	\epsilon_\mathrm{opt}(\partial\partial' f_\mathrm{Q})\cong&\,\left[\dfrac{2304d}{\MEAN{(\partial\partial' f_{\mathrm{Q}})^2}{}N_\mathrm{T}(d+1)}\right]^{1/8}\,,
\end{align}
along with the optimized MSEs:
\begin{align}
	\MSE{\mathrm{FD,opt}}{\partial f_\mathrm{Q}}=&\,\left[\frac{3\,d^2 \MEAN{(\partial f_\mathrm{Q})^2}{}}{16\,N^2_{\mathrm{T}}\,(d+1)^2}\right]^{1/3}\,,\nonumber\\
	\MSE{\mathrm{FD,opt}}{\partial\partial f_\mathrm{Q}}=&\,\sqrt{\frac{d\MEAN{(\partial\partial f_\mathrm{Q})^2}{}}{2\,N_{\mathrm{T}}(d+1)}}\,,\nonumber\\
	\MSE{\mathrm{FD,opt}}{\partial\partial' f_\mathrm{Q}}=&\,\dfrac{2}{3}\sqrt{\frac{d\MEAN{(\partial\partial' f_\mathrm{Q})^2}{}}{N_{\mathrm{T}}(d+1)}}\,.
	\label{eq:FDMSEopt_gen}
\end{align}

The final task is then to substitute the correct expressions of $\MEAN{\partial f_\mathrm{Q,k})^2}{}$, $\MEAN{\partial\partial f_\mathrm{Q,k})^2}{}$ and $\MEAN{\partial\partial' f_\mathrm{Q,k})^2}{}$ that are applicable to the relevant case in point as listed in Fig.~\ref{fig:cases}. Enjoying the fruits of our labor in Appendices~\ref{app:avgdf} through \ref{app:avgddf}, summarized in Tab.~\ref{tab:avg_res}, we observe that only {\bf Case~I} supplies exact analytical expressions of these averages, whereas all other cases provide only upper bounds. Optimally-tuned estimators, therefore, refer to either those that minimizes the exact expression of two-design-averaged MSEs in {\bf Case~I}, or MSE upper bounds in all other cases.

Analytical formulations of approximately optimal FD estimators for \emph{any} case require substitutions of the answers from Tab.~\ref{tab:avg_res}, which may be done if so desired. For benchmarking with the PS strategy, however, all one needs is to recognize that Tab.~\ref{tab:avg_res} implies that $\MEAN{\partial f_\mathrm{Q,k}}{},\MEAN{\partial\partial f_\mathrm{Q,k}}{},\MEAN{\partial\partial' f_\mathrm{Q,k}}{}\leq O(1/d)$ for large $d$, so that
\begin{align}
	&\,\MSE{\mathrm{FD,opt}}{\partial f_\mathrm{Q}}\lesssim O\left(\dfrac{1}{N^{2/3}_{\mathrm{T}}d^{1/3}}\right)\,,\nonumber\\
	&\,\MSE{\mathrm{FD,opt}}{\partial\partial f_\mathrm{Q}},\MSE{\mathrm{FD,opt}}{\partial\partial' f_\mathrm{Q}}\lesssim O\left(\dfrac{1}{N^{1/2}_{\mathrm{T}}d^{1/2}}\right),
\end{align}
all scales exponentially with $n$, which are compatible with the barren-plateau phenomenon that also commensurately scales all gradient- and Hessian-component squared-magnitudes exponentially with $n$.

\subsection{Optimally-tuned GD estimators for general cases}
\label{app:GDallcase}

The generalization of FD estimators as defined by \eqref{eq:GD1_est} and \eqref{eq:GD2_est}, carries the same basic functional structure $\mathrm{MSE}_\mathrm{GD}=\rvec{c}^\top\dyadic{M}\,\rvec{c}$ in their MSEs in \eqref{eq:GD_MSE} and \eqref{eq:GD_MSE2}. The only additional step one needs to perform is the minimization over normalized $\rvec{c}$. To do this, we first introduce the Lagrange function
\begin{equation}
	\mathcal{D}=\rvec{c}^\top\dyadic{M}\,\rvec{c}-2\,\lambda\,(\rvec{c}^\top\onevec-1)
\end{equation}
that is to be optimized, where $\lambda$ is the Lagrange multiplier that takes care of the normalization constraint, and the factor of 2 in front of $\lambda$ is introduced for convenience that shall become clear very soon. An \emph{arbitrary} variation of $\mathcal{D}$ over $\rvec{c}$ gives
\begin{equation}
	\updelta\mathcal{D}=\updelta\rvec{c}^\top\,\dyadic{M}\,\rvec{c}+\rvec{c}^\top\,\dyadic{M}\,\updelta\rvec{c}-\lambda\,(\updelta\rvec{c}^\top\onevec+\onevec^\top\updelta\rvec{c})\,.
\end{equation} 
A minimization of $\mathcal{D}$, akin to the constrained minimization of $\mathrm{MSE}_\mathrm{GD}$, is done by setting $\updelta\mathcal{D}=0$, resulting in $\dyadic{M}\,\rvec{c}=\lambda\onevec$. As $\dyadic{M}$ is invertible, solving $\lambda$ yields the extremal equation
\begin{equation}
	\rvec{c}=\dfrac{\dyadic{M}^{-1}\onevec}{\onevec^\top\dyadic{M}^{-1}\onevec}\,.
	\label{eq:copt}
\end{equation}
In other words, the optimal $\mathrm{MSE}_\mathrm{GD}$s may be obtained by first substituting $\rvec{c}$ $\mathrm{MSE}_\mathrm{GD}$ with the optimal one defined in \eqref{eq:copt}, and, next, minimizing the result over $\epsilon$.

Just like for FD estimators, this second minimization over $\epsilon$ can be done numerically, the results of which are used in Figs.~\ref{fig:MSEGD} and \ref{fig:perf_8qbits}. To benchmark GD estimators with the PS~ones, we may again resort to looking at upper bounds of $\mathrm{MSE}_\mathrm{GD}$. Thankfully, the Cauchy--Schwarz inequality remains our dearest friend for this task, awarding us with $\vphantom{{W^W}^W}(\onevec^\top\onevec)^2\leq\onevec^\top\dyadic{M}\onevec\onevec^\top\dyadic{M}^{-1}\onevec$, and consequently $\mathrm{MSE}_{\mathrm{GD,opt}}\leq J^{-2}\min_{\epsilon>0}\onevec^\top\dyadic{M}\onevec$.

\begin{table*}
	\begin{tabular}[t]{rrcccccccc}
		$k$&$h_k$ & $O_{k1}$ & $O_{k2}$ & $O_{k3}$ & $O_{k4}$ & $O_{k5}$ & $O_{k6}$ & $O_{k7}$ & $O_{k8}$ \\
		\hline\hline							
		{\bf1}&$-0.180625859$ & $1$ & $1$ & $1$ & $1$ & $1$ & $1$ & $Z$ & $1$\\							
		{\bf2}&$-0.180625859$ & $1$ & $1$ & $1$ & $1$ & $1$ & $1$ & $1$ & $Z$\\														
		{\bf3}&$-0.159587991$ & $1$ & $1$ & $1$ & $1$ & $Z$ & $1$ & $1$ & $1$\\														
		{\bf4}&$-0.159587991$ & $1$ & $1$ & $1$ & $1$ & $1$ & $Z$ & $1$ & $1$\\							
		{\bf5}&$0.174193924$	& $1$ & $1$ & $Z$ & $1$ & $1$ & $1$ & $1$ & $1$\\							
		{\bf6}&$0.174193924$	& $1$ & $1$ & $1$ & $Z$ & $1$ & $1$ & $1$ & $1$\\							
		{\bf7}&$0.227570968$	& $Z$ & $1$ & $1$ & $1$ & $1$ & $1$ & $1$ & $1$\\							
		{\bf8}&$0.227570968$	& $1$ & $Z$ & $1$ & $1$ & $1$ & $1$ & $1$ & $1$\\							
		{\bf9}&$0.112704888$	& $1$ & $1$ & $1$ & $1$ & $Z$ & $1$ & $Z$ & $1$\\						
		{\bf10}&$0.112704888$	& $1$ & $1$ & $1$ & $1$ & $1$ & $Z$ & $1$ & $Z$\\						
		{\bf11}&$0.119520678$	& $Z$ & $1$ & $1$ & $1$ & $Z$ & $1$ & $1$ & $1$\\						
		{\bf12}&$0.119520678$	& $1$ & $Z$ & $1$ & $1$ & $1$ & $Z$ & $1$ & $1$\\						
		{\bf13}&$0.134010069$	& $Z$ & $1$ & $1$ & $1$ & $1$ & $1$ & $Z$ & $1$\\						
		{\bf14}&$0.134010069$	& $1$ & $Z$ & $1$ & $1$ & $1$ & $1$ & $1$ & $Z$\\						
		{\bf15}&$0.137351346$	& $Z$ & $1$ & $1$ & $1$ & $1$ & $Z$ & $1$ & $1$\\						
		{\bf16}&$0.137351346$	& $1$ & $Z$ & $1$ & $1$ & $Z$ & $1$ & $1$ & $1$\\						
		{\bf17}&$0.137670707$	& $1$ & $1$ & $Z$ & $1$ & $Z$ & $1$ & $1$ & $1$\\						
		{\bf18}&$0.137670707$	& $1$ & $1$ & $1$ & $Z$ & $1$ & $Z$ & $1$ & $1$\\						
		{\bf19}&$0.141387333$	& $1$ & $1$ & $1$ & $1$ & $Z$ & $1$ & $1$ & $Z$\\						
		{\bf20}&$0.141387333$	& $1$ & $1$ & $1$ & $1$ & $1$ & $Z$ & $Z$ & $1$\\						
		{\bf21}&$0.147230746$	& $1$ & $1$ & $Z$ & $1$ & $1$ & $Z$ & $1$ & $1$\\						
		{\bf22}&$0.147230746$	& $1$ & $1$ & $1$ & $Z$ & $Z$ & $1$ & $1$ & $1$\\						
		{\bf23}&$0.149265817$	& $1$ & $1$ & $1$ & $1$ & $Z$ & $Z$ & $1$ & $1$\\						
		{\bf24}&$0.149731063$	& $1$ & $1$ & $Z$ & $1$ & $1$ & $1$ & $Z$ & $1$\\						
		{\bf25}&$0.149731063$	& $1$ & $1$ & $1$ & $Z$ & $1$ & $1$ & $1$ & $Z$\\						
		{\bf26}&$0.151377428$	& $Z$ & $1$ & $1$ & $1$ & $1$ & $1$ & $1$ & $Z$\\						
		{\bf27}&$0.151377428$	& $1$ & $Z$ & $1$ & $1$ & $1$ & $1$ & $Z$ & $1$\\						
		{\bf28}&$0.154354359$	& $1$ & $1$ & $1$ & $1$ & $1$ & $1$ & $Z$ & $Z$\\						
		{\bf29}&$0.155818816$	& $1$ & $1$ & $Z$ & $1$ & $1$ & $1$ & $1$ & $Z$\\						
		{\bf30}&$0.155818816$	& $1$ & $1$ & $1$ & $Z$ & $1$ & $1$ & $Z$ & $1$\\						
		{\bf31}&$0.167560719$	& $Z$ & $1$ & $Z$ & $1$ & $1$ & $1$ & $1$ & $1$\\						
		{\bf32}&$0.167560719$	& $1$ & $Z$ & $1$ & $Z$ & $1$ & $1$ & $1$ & $1$\\						
		{\bf33}&$0.181433521$	& $Z$ & $1$ & $1$ & $Z$ & $1$ & $1$ & $1$ & $1$\\						
		{\bf34}&$0.181433521$	& $1$ & $Z$ & $Z$ & $1$ & $1$ & $1$ & $1$ & $1$\\						
		{\bf35}&$0.19391146$	& $Z$ & $Z$ & $1$ & $1$ & $1$ & $1$ & $1$ & $1$\\						
		{\bf36}&$0.220039773$ & $1$ & $1$ & $Z$ & $Z$ & $1$ & $1$ & $1$ & $1$\\						
		{\bf37}&$-0.028682446$ & $1$ & $1$ & $1$ & $1$ & $Y$ & $Y$ & $X$ & $X$\\
		{\bf38}&$-0.028682446$ & $1$ & $1$ & $1$ & $1$ & $X$ & $X$ & $Y$ & $Y$\\
		{\bf39}&$-0.017830668$ & $Y$ & $Y$ & $1$ & $1$ & $X$ & $X$ & $1$ & $1$\\
		{\bf40}&$-0.017830668$ & $X$ & $X$ & $1$ & $1$ & $Y$ & $Y$ & $1$ & $1$\\
		{\bf41}&$-0.01736736$ & $Y$ & $Y$ & $1$ & $1$ & $1$ & $1$ & $X$ & $X$\\
		{\bf42}&$-0.01736736$ & $X$ & $X$ & $1$ & $1$ & $1$ & $1$ & $Y$ & $Y$\\
		{\bf43}&$-0.013872802$ & $Y$ & $Y$ & $X$ & $X$ & $1$ & $1$ & $1$ & $1$\\
		{\bf44}&$-0.013872802$ & $X$ & $X$ & $Y$ & $Y$ & $1$ & $1$ & $1$ & $1$\\
		{\bf45}&$-0.00956004$	& $1$ & $1$ & $Y$ & $Y$ & $X$ & $X$ & $1$ & $1$\\
		{\bf46}&$-0.00956004$ & $1$ & $1$ & $X$ & $X$ & $Y$ & $Y$ & $1$ & $1$\\
		{\bf47}&$-0.006087753$ & $1$ & $1$ & $Y$ & $Y$ & $1$ & $1$ & $X$ & $X$\\
		{\bf48}&$-0.006087753$ & $1$ & $1$ & $X$ & $X$ & $1$ & $1$ & $Y$ & $Y$
	\end{tabular}\quad
	\begin{tabular}[t]{rrcccccccc}
		$k$ & $h_k$ & $O_{k1}$ & $O_{k2}$ & $O_{k3}$ & $O_{k4}$ & $O_{k5}$ & $O_{k6}$ & $O_{k7}$ & $O_{k8}$ \\
		\hline\hline	
		{\bf49}&$0.006087753$ & $1$ & $1$ & $Y$ & $X$ & $1$ & $1$ & $X$ & $Y$\\
		{\bf50}&$0.006087753$ & $1$ & $1$ & $X$ & $Y$ & $1$ & $1$ & $Y$ & $X$\\
		{\bf51}&$0.00956004$ & $1$ & $1$ & $Y$ & $X$ & $X$ & $Y$ & $1$ & $1$\\
		{\bf52}&$0.00956004$ & $1$ & $1$ & $X$ & $Y$ & $Y$ & $X$ & $1$ & $1$\\
		{\bf53}&$0.01130811$ & $1$ & $Y$ & $Z$ & $Z$ & $1$ & $Y$ & $1$ & $1$\\
		{\bf54}&$0.01130811$ & $1$ & $X$ & $Z$ & $Z$ & $1$ & $X$ & $1$ & $1$\\
		{\bf55}&$0.013872802$ & $Y$ & $X$ & $X$ & $Y$ & $1$ & $1$ & $1$ & $1$\\
		{\bf56}&$0.013872802$ & $X$ & $Y$ & $Y$ & $X$ & $1$ & $1$ & $1$ & $1$\\
		{\bf57}&$0.01736736$ & $Y$ & $X$ & $1$ & $1$ & $1$ & $1$ & $X$ & $Y$\\
		{\bf58}&$0.01736736$ & $X$ & $Y$ & $1$ & $1$ & $1$ & $1$ & $Y$ & $X$\\
		{\bf59}&$0.017830668$ & $Y$ & $X$ & $1$ & $1$ & $X$ & $Y$ & $1$ & $1$\\
		{\bf60}&$0.017830668$ & $X$ & $Y$ & $1$ & $1$ & $Y$ & $X$ & $1$ & $1$\\
		{\bf61}&$0.028682446$ & $1$ & $1$ & $1$ & $1$ & $Y$ & $X$ & $X$ & $Y$\\
		{\bf62}&$0.028682446$ & $1$ & $1$ & $1$ & $1$ & $X$ & $Y$ & $Y$ & $X$\\
		{\bf63}&$0.029818179$ & $Y$ & $Z$ & $Z$ & $1$ & $Y$ & $1$ & $1$ & $1$\\
		{\bf64}&$0.029818179$	& $X$ & $Z$ & $Z$ & $1$ & $X$ & $1$ & $1$ & $1$\\
		{\bf65}&$0.029818179$ & $1$ & $Y$ & $1$ & $Z$ & $Z$ & $Y$ & $1$ & $1$\\	
		{\bf66}&$0.029818179$ & $1$ & $X$ & $1$ & $Z$ & $Z$ & $X$ & $1$ & $1$\\
		{\bf67}&$0.030109333$	& $Y$ & $Z$ & $1$ & $Z$ & $Y$ & $1$ & $1$ & $1$\\
		{\bf68}&$0.030109333$ & $X$ & $Z$ & $1$ & $Z$ & $X$ & $1$ & $1$ & $1$\\
		{\bf69}&$0.030109333$ & $1$ & $Y$ & $Z$ & $1$ & $Z$ & $Y$ & $1$ & $1$\\
		{\bf70}&$0.030109333$ & $1$ & $X$ & $Z$ & $1$ & $Z$ & $X$ & $1$ & $1$\\
		{\bf71}&$0.030791132$ & $Y$ & $1$ & $Z$ & $Z$ & $Y$ & $1$ & $1$ & $1$\\
		{\bf72}&$0.030791132$ & $X$ & $1$ & $Z$ & $Z$ & $X$ & $1$ & $1$ & $1$\\
		{\bf73}&$0.043763244$ & $Y$ & $Z$ & $Z$ & $Z$ & $Y$ & $1$ & $1$ & $1$\\
		{\bf74}&$0.043763244$ & $X$ & $Z$ & $Z$ & $Z$ & $X$ & $1$ & $1$ & $1$\\
		{\bf75}&$0.043763244$ & $1$ & $Y$ & $Z$ & $Z$ & $Z$ & $Y$ & $1$ & $1$\\
		{\bf76}&$0.043763244$ & $1$ & $X$ & $Z$ & $Z$ & $Z$ & $X$ & $1$ & $1$\\
		{\bf77}&$-0.0145648$ & $1$ & $Y$ & $Z$ & $Z$ & $X$ & $1$ & $X$ & $Y$\\
		{\bf78}&$-0.0145648$ & $1$ & $Y$ & $Z$ & $Z$ & $Y$ & $1$ & $Y$ & $Y$\\
		{\bf79}&$-0.0145648$ & $1$ & $X$ & $Z$ & $Z$ & $X$ & $1$ & $X$ & $X$\\
		{\bf80}&$-0.0145648$ & $1$ & $X$ & $Z$ & $Z$ & $Y$ & $1$ & $Y$ & $X$\\
		{\bf81}&$0.010541633$ & $Y$ & $Z$ & $Z$ & $Z$ & $Y$ & $1$ & $1$ & $Z$\\
		{\bf82}&$0.010541633$ & $X$ & $Z$ & $Z$ & $Z$ & $X$ & $1$ & $1$ & $Z$\\
		{\bf83}&$0.010541633$ & $1$ & $Y$ & $Z$ & $Z$ & $Z$ & $Y$ & $Z$ & $1$\\
		{\bf84}&$0.010541633$ & $1$ & $X$ & $Z$ & $Z$ & $Z$ & $X$ & $Z$ & $1$\\
		{\bf85}&$0.01130811$ & $Y$ & $Z$ & $Z$ & $Z$ & $Y$ & $Z$ & $1$ & $1$\\
		{\bf86}&$0.01130811$ & $X$ & $Z$ & $Z$ & $Z$ & $X$ & $Z$ & $1$ & $1$\\
		{\bf87}&$0.025106432$ & $Y$ & $Z$ & $Z$ & $Z$ & $Y$ & $1$ & $Z$ & $1$\\
		{\bf88}&$0.025106432$ & $X$ & $Z$ & $Z$ & $Z$ & $X$ & $1$ & $Z$ & $1$\\
		{\bf89}&$0.025106432$ & $1$ & $Y$ & $Z$ & $Z$ & $Z$ & $Y$ & $1$ & $Z$\\
		{\bf90}&$0.025106432$ & $1$ & $X$ & $Z$ & $Z$ & $Z$ & $X$ & $1$ & $Z$\\
		{\bf91}&$0.030791132$ & $Z$ & $Y$ & $Z$ & $Z$ & $Z$ & $Y$ & $1$ & $1$\\
		{\bf92}&$0.030791132$ & $Z$ & $X$ & $Z$ & $Z$ & $Z$ & $X$ & $1$ & $1$\\
		{\bf93}&$-0.0145648$ & $Y$ & $Z$ & $Z$ & $Z$ & $Z$ & $Y$ & $X$ & $X$\\
		{\bf94}&$-0.0145648$ & $X$ & $Z$ & $Z$ & $Z$ & $Z$ & $X$ & $Y$ & $Y$\\
		{\bf95}&$0.0145648$ & $Y$ & $Z$ & $Z$ & $Z$ & $Z$ & $X$ & $X$ & $Y$\\
		{\bf96}&$0.0145648$ & $X$ & $Z$ & $Z$ & $Z$ & $Z$ & $Y$ & $Y$ & $X$		
	\end{tabular}
	\caption{\label{tab:water}The complete sheet of all 96 multiqubit Pauli components and their respective weights (of magnitudes larger than $10^{-3}$) that constitute the observable $O$ describing a neutral water molecule in the minimal basis set \texttt{sto-3g}, where electronic excitations are restricted to four active orbitals. The two hydrogen~(H) atoms and one oxygen~(O) atom are geometrically positioned according to the respective spatial ($x,y,z$) coordinates---H:~$(-0.0399, -0.0038, 0.0)$, O:~$(1.5780, 0.8540, 0.0)$, H:~$(2.7909, -0.5159, 0.0)$. All coefficients are generated using the \texttt{pennylane.qchem} Python package.~\footnote{Visit \url{https://pennylane.ai/qml/demos/tutorial_quantum_chemistry.html}}}
\end{table*}

We now take advantage of the fact that the $J$-dimensional $\dyadic{M}$s, which are more precisely listed in \eqref{eq:GD_MSE}, give rise to \mbox{$M\equiv\onevec^\top\dyadic{M}\,\onevec$} that are multidimensional analogs of the right-hand sides in \eqref{eq:FD_MSE_gen}, namely
\begin{align}
	M_{\partial f_\mathrm{Q}}\cong&\,\dfrac{4Jd\,\HARM{J,2}}{N_\mathrm{T}(d+1)\epsilon^2}\nonumber\\
	&\,+\epsilon^4\MEAN{(\partial f_{\mathrm{Q,k}})^2}{}\frac{J^2 (J+1)^2 (2 J+1)^2}{20736}\,,\nonumber\\
	M_{\partial\partial f_\mathrm{Q}}\cong&\,\dfrac{(2J+1)d}{N_\mathrm{T}(d+1)\epsilon^4}(4\HARM{J,2}^2+2\HARM{J,4})\nonumber\\
	&\,+\epsilon^4\MEAN{(\partial\partial f_{\mathrm{Q,k}})^2}{}\frac{J^2 (J+1)^2 (2 J+1)^2}{5184}\,,\nonumber\\
	M_{\partial\partial' f_\mathrm{Q}}\cong&\,\dfrac{16Jd\,\HARM{J,4}}{N_\mathrm{T}(d+1)\epsilon^4}\nonumber\\
	&\,+\epsilon^4\MEAN{(\partial\partial' f_{\mathrm{Q,k}})^2}{}\frac{J^2 (J+1)^2 (2 J+1)^2}{5184}\,,
	\label{eq:GD_MSE_gen}
\end{align}
for $\epsilon\ll1$, where $\HARM{j,k}=\sum^j_{m=1}m^{-k}$ is the generalized harmonic number. Recalling the results in \eqref{eq:min_twostructs}, we write down the optimized upper bounds
\begin{widetext}
\begin{align}
	\MSE{\mathrm{GD,opt}}{\partial f_\mathrm{Q}}\lesssim&\,\dfrac{1}{J^2}\left[\dfrac{\MEAN{(\partial f)^2}{} d^2 J^4 (1 + J)^2 (1 + 2 J)^2\, \HARM{J,2}^2}{192 N^2_\mathrm{T}(1 + d)^2}\right]^{1/3}\,,\nonumber\\
	\MSE{\mathrm{GD,opt}}{\partial\partial f_\mathrm{Q}}\lesssim&\,\dfrac{1}{36J^2} \sqrt{\frac{ \MEAN{(\partial\partial f)^2}{}d J^2 (J+1)^2 (2 J+1)^3 \left(4 \HARM{J,2}^2+2 \HARM{J,4}\right)}{N_\mathrm{T}(d+1)}}\,,\nonumber\\
	\MSE{\mathrm{GD,opt}}{\partial\partial' f_\mathrm{Q}}\lesssim&\,\frac{1}{9J^2} \sqrt{\frac{\MEAN{(\partial\partial' f)^2}{} d J^3 (1 + J)^2 (1 + 2 J)^2\, \HARM{J,4}}{N_\mathrm{T}(d+1)}}\,,
	\label{eq:GDMSEopt_gen}
\end{align}
\end{widetext}
all of which reduce to the corresponding \emph{equalities} in \eqref{eq:FDMSEopt_gen} for $J=1$ as $\HARM{1,k}=1$. Unsurprisingly, for large $d$,
\begin{align}
	&\,\MSE{\mathrm{GD,opt}}{\partial f_\mathrm{Q}}\lesssim O\left(\dfrac{1}{N^{2/3}_{\mathrm{T}}d^{1/3}}\right)\,,\nonumber\\
	&\,\MSE{\mathrm{GD,opt}}{\partial\partial f_\mathrm{Q}},\MSE{\mathrm{GD,opt}}{\partial\partial' f_\mathrm{Q}}\lesssim O\left(\dfrac{1}{N^{1/2}_{\mathrm{T}}d^{1/2}}\right)\!.
\end{align}
These inequalities are sufficient to again show that the optimal MSEs for the GD estimators scale with the respective component squared-magnitudes.

\subsection{Optimally-tuned SPS estimators for general cases}
\label{app:SPSallcase}

For general cases, the $\MSE{\mathrm{SPS}}{\,\cdot\,}$ expressions read
\begin{align}
	\MSE{\mathrm{SPS}}{\partial f_\mathrm{Q}}=&\,\dfrac{d\lambda^2}{N_\mathrm{T}(d+1)}+\MEAN{(\partial f_{\mathrm{Q}})^2}{}(1-\lambda)^2\,,\nonumber\\
	\MSE{\mathrm{SPS}}{\partial\partial f_\mathrm{Q}}=&\,\dfrac{9d\lambda^2}{8N_\mathrm{T}(d+1)}+\MEAN{(\partial\partial f_{\mathrm{Q}})^2}{}(1-\lambda)^2\,,\nonumber\\
	&\,\qquad\qquad{\text{(diagonal Hessian components)}}\nonumber\\
	\MSE{\mathrm{SPS}}{\partial\partial' f_\mathrm{Q}}=&\,\dfrac{d\lambda^2}{N_\mathrm{T}(d+1)}+\MEAN{(\partial\partial' f_{\mathrm{Q}})^2}{}(1-\lambda)^2\,,\nonumber\\
	&\,\qquad\quad\!\!\!{\text{(off-diagonal Hessian components)}}
	\label{eq:SPS_MSE_gen}
\end{align}
where $s=\pi/2$. After optimizing the scaling prefactors,
\begin{align}
	\MSE{\mathrm{SPS,opt}}{\partial f_\mathrm{Q}}=&\,\dfrac{d\MEAN{(\partial f_{\mathrm{Q}})^2}{}}{d + (d+1)\MEAN{(\partial f_{\mathrm{Q}})^2}{}N_\mathrm{T}}\,,\nonumber\\
	\MSE{\mathrm{SPS,opt}}{\partial\partial f_\mathrm{Q}}=&\,\dfrac{9d\MEAN{(\partial\partial f_{\mathrm{Q}})^2}{}}{9d + 8(d+1)\MEAN{(\partial\partial f_{\mathrm{Q}})^2}{}N_\mathrm{T}}\,,\nonumber\\
	&\,\qquad\,{\text{(diagonal Hessian components)}}\nonumber\\
	\MSE{\mathrm{SPS,opt}}{\partial\partial' f_\mathrm{Q}}=&\,\dfrac{d\MEAN{(\partial\partial' f_{\mathrm{Q}})^2}{}}{d + (d+1)\MEAN{(\partial\partial' f_{\mathrm{Q}})^2}{}N_\mathrm{T}}\,.\nonumber\\
	&\,\quad\!\!{\text{(off-diagonal Hessian components)}}
	\label{eq:SPS_MSE_opt_gen}
\end{align}

It is straightforward to verify that $\MSE{\mathrm{SPS,opt}}{\partial f_\mathrm{Q}}$, $\MSE{\mathrm{SPS,opt}}{\partial\partial f_\mathrm{Q}}$ and $\MSE{\mathrm{SPS,opt}}{\partial\partial' f_\mathrm{Q}}$ are respectively monotonically increasing in $\MEAN{(\partial f_{\mathrm{Q}})^2}{}$, $\MEAN{(\partial\partial f_{\mathrm{Q}})^2}{}$ and $\MEAN{(\partial\partial' f_{\mathrm{Q}})^2}{}$, since the derivatives 
\begin{align}
	\dfrac{\partial\MSE{\mathrm{SPS,opt}}{\partial f_\mathrm{Q}}}{\partial\MEAN{(\partial f_{\mathrm{Q}})^2}{}}=&\,\dfrac{d^2}{[d + (d+1)\MEAN{(\partial f_{\mathrm{Q}})^2}{} N_\mathrm{T}]^2}\,,\nonumber\\
	\dfrac{\partial\MSE{\mathrm{SPS,opt}}{\partial\partial f_\mathrm{Q}}}{\partial\MEAN{(\partial\partial f_{\mathrm{Q}})^2}{}}=&\,\dfrac{81d^2}{[9d + 8(d+1)\MEAN{(\partial\partial f_{\mathrm{Q}})^2}{} N_\mathrm{T}]^2}\,,\nonumber\\
	\dfrac{\partial\MSE{\mathrm{SPS,opt}}{\partial\partial' f_\mathrm{Q}}}{\partial\MEAN{(\partial\partial' f_{\mathrm{Q}})^2}{}}=&\,\dfrac{d^2}{[d + (d+1)\MEAN{(\partial\partial' f_{\mathrm{Q}})^2}{} N_\mathrm{T}]^2}\,,
\end{align}
are all nonnegative. It then follows from Tab.~\ref{tab:avg_res} that the SPS MSEs are always smaller than the PS MSEs as $\MEAN{(\partial f_{\mathrm{Q}})^2}{}$, $\MEAN{(\partial\partial f_{\mathrm{Q}})^2}{}$ and $\MEAN{(\partial\partial' f_{\mathrm{Q}})^2}{}$ are all less than one. Based on~\eqref{eq:SPS_MSE_opt_gen}, in the limit $N_\mathrm{T}\gg d$, we find that $\MSE{\mathrm{SPS,opt}}{\,\cdot\,}\rightarrow\MSE{\mathrm{PS}}{\,\cdot\,}$\,.

\section{Simplified electronic description of a water molecule}
\label{app:water}

In the quantum-eigensolver scenario, the observable $O=H-h_01$, with $h_0=\tr{H}/d$, is defined as a trace-subtracted Hamilton operator $H$ that describes the dynamics of electrons in a water molecule. Under the Hartree--Fock approximation~\cite{Seeger:1977self}, every electron in the molecule is treated as an independent particle that experiences both the Coulomb potential from the nuclei and a mean field generated by all other electrons. The results in Fig.~\ref{fig:perf_8qbits}(b) are generated by imposing an additional restriction on the electronic excitation to four active orbitals. An application of the Jordan--Wigner transformation turns the resulting Hartree--Fock Hamilton operator into a linear combination of multiqubit Pauli operators weighted by coefficients listed in Tab.~\ref{tab:water}.

\end{document}